\documentclass{article}

\usepackage{hyperref}

\usepackage[accepted]{icml2026}

\usepackage{amsmath,amsfonts,bm}

\def\eqref#1{equation~\ref{#1}}

\def\1{\bm{1}}

\DeclareMathAlphabet{\mathsfit}{\encodingdefault}{\sfdefault}{m}{sl}
\SetMathAlphabet{\mathsfit}{bold}{\encodingdefault}{\sfdefault}{bx}{n}

\usepackage{hyperref}
\usepackage{url}
\usepackage{cleveref}
\usepackage{booktabs}
\usepackage{graphicx}
\usepackage{xspace}
\usepackage{array}
\usepackage{enumitem}
\usepackage{tabularx}
\usepackage{booktabs}
\usepackage{makecell}
\usepackage{graphicx}
\usepackage{subcaption}
\usepackage{tikz}

\usepackage{amsmath}
\usepackage{amssymb}
\usepackage{mathtools}
\usepackage{amsthm}

\newcommand{\ourmodel}{ProtDBench\xspace}

\theoremstyle{plain}

\theoremstyle{definition}

\theoremstyle{remark}

\usepackage{multirow} 

\usepackage[textsize=tiny]{todonotes}

\icmltitlerunning{ProtDBench: A Unified Benchmark of Protein Binder Design and Evaluation
}

\begin{document}

\twocolumn[
  \icmltitle{ProtDBench: A Unified Benchmark of Protein Binder Design and Evaluation
}



  \icmlsetsymbol{equal}{*}

  \begin{icmlauthorlist}
    \icmlauthor{Cong Liu}{amlab,bdpast}
    \icmlauthor{Milong Ren}{bdpast}
    \icmlauthor{Jiaqi Guan}{bd}
    \icmlauthor{Chengyue Gong}{bd}
    \icmlauthor{Jinyuan Sun}{bdpast}
    \icmlauthor{Xinshi Chen}{bdpast}
    \icmlauthor{Wenzhi Xiao}{bdpast}
  \end{icmlauthorlist}

  \icmlaffiliation{amlab}{AMLab, AI4Science Lab, University of Amsterdam, Amsterdam, The Netherlands}
  \icmlaffiliation{bd}{ByteDance Seed}
  \icmlaffiliation{bdpast}{Work done at ByteDance Seed}

  \icmlcorrespondingauthor{Wenzhi Xiao}{xwzded00@163.com}

  \icmlkeywords{Machine Learning, ICML}

  \vskip 0.3in
]

\printAffiliationsAndNotice{}  

\begin{abstract}
Recent advances in \emph{de novo} protein binder design have enabled increasing experimental validation, yet reported \emph{in silico} metrics remain difficult to interpret or compare across studies due to non-standardized evaluation protocols. We introduce \textbf{ProtDBench}\footnote{Code: \url{https://github.com/congliuUvA/ProtDBench}}, a standardized and throughput-aware evaluation framework for protein binder design. ProtDBench defines unified benchmark tasks, evaluation protocols, and success criteria, enabling systematic analysis of how evaluation design influences observed performance. Using a large wet-lab annotated dataset, we analyze commonly used structure prediction models as evaluation verifiers, revealing substantial verifier-dependent bias and limited agreement under identical filtering protocols. We then benchmark representative open-source generative binder design methods across ten diverse protein targets under a fixed evaluation protocol. Beyond per-sequence success rates, ProtDBench incorporates throughput-aware metrics based on a fixed 24-hour budget, as well as cluster-level success criteria to account for structural diversity. Together, these results expose systematic differences induced by filtering rules, success definitions, and throughput-aware evaluation between computational efficiency, success rate, and structural diversity. Overall, ProtDBench provides a fair and reproducible evaluation pipeline that supports systematic and controlled comparison of protein binder design methods under realistic evaluation settings.

\end{abstract}

\section{Introduction}

Protein binder design, the task of engineering proteins that bind with high affinity and specificity to molecular targets, 
is a central challenge in therapeutic development, diagnostics, and synthetic biology. Traditional approaches rely on physics-based modeling or extensive experimental screening, which are often slow, costly, and labor-intensive. 
Recent advances in biomolecular structure prediction and generative modeling have substantially accelerated \emph{de novo} protein binder design.
Building on accurate structure predictors, modern approaches can rapidly generate large numbers of candidate binders, enabling both diffusion-based and hallucination-based design paradigms \citep[e.g.,][]{chu2023allatom, pacesa2025one, cho2025boltzdesign1, Stark2025BoltzGen, watson2023novo, butcher2025novo, ren2025pxdesign, zhang2025odesignworldmodelbiomolecular}.

Despite rapid progress in protein binder generation, \emph{evaluation} remains a critical and underdeveloped component of current design pipelines.
In practice, success is assessed almost exclusively through \emph{in silico} filtering using confidence scores derived from structure prediction models~\citep{watson2023novo,Stark2025BoltzGen}.
However, this evaluation paradigm suffers from several limitations:

\begin{itemize}[left=6pt]
    \item \textbf{Reliability of structure-based filters}
    Confidence metrics (e.g., pLDDT, ipTM, and ipAE) are widely used as proxies for binding success, yet their correlation with experimental outcomes has only been validated on limited datasets and remains poorly understood across targets and design regimes.

    \item \textbf{Standardization of benchmarking protocols}
    Evaluation practices vary substantially in target selection, hotspot definitions, filtering thresholds, verifier choice, and computational budgets, making fair comparison between generative methods difficult~\citep[e.g.,][]{watson2023novo,pacesa2025one, Stark2025BoltzGen}.

    \item \textbf{Awareness of throughput related metrics}
    Existing reported performance is often reduced to per-sample success rates, obscuring important trade-offs between effectiveness and efficiency, such as generation throughput,
    diversity of viable binders, 
    and resource-constrained performance.

    \item \textbf{Reproducibility of filtering pipelines}
    Many evaluation pipelines rely on implicit design choices or filtering configurations, limiting reproducibility and hindering systematic analysis of how evaluation decisions influence reported results \citep{zambaldi2024novo, Stark2025BoltzGen}.
\end{itemize}

To address these challenges, we propose \textbf{\ourmodel}, a unified and throughput-aware evaluation framework for \emph{de novo} protein binder design.
\ourmodel is built around the observation that evaluation choices, 
such as the structure prediction verifier, filtering strategy, and computational budget, substantially shape reported performance, yet are rarely made explicit or analyzed systematically.
By grounding evaluation in wet-lab annotated data, \ourmodel enables principled assessment of the reliability and bias of structure-based confidence metrics.
At the same time, \ourmodel standardizes benchmark targets, verifier configurations, and filtering protocols, allowing fair and reproducible comparison across generative methods.
Beyond per-sample success rates, \ourmodel incorporates throughput and diversity-aware metrics that better reflect practical design constraints.

We instantiate \textbf{\ourmodel} through a comprehensive two-pillar study. 
Firstly, we conduct large-scale retrospective analyses on a large-scale wet-lab annotated dataset, Cao dataset \citep{cao2022design}, to quantify the predictive power and complementarity of commonly used structure prediction–based verifiers. 
Secondly, under a standardized evaluation protocol, we benchmark a wide range of representative open-source generative methods across diverse protein targets, analyzing not only success rates but also cluster-level diversity, structural consistency, and computational throughput.

\section{Evaluation Framework: \ourmodel}
\label{sec:framework}

\textbf{\ourmodel} (\Cref{alg:protdbench}) is a unified evaluation framework for \emph{de novo} protein binder design.
Rather than treating evaluation as a fixed component of a design pipeline, \ourmodel explicitly formalizes evaluation as a configurable process that operates on generated candidates and is independent of the generative mechanism itself. 
This formulation enables controlled, reproducible comparisons and facilitates systematic analysis of how evaluation choices influence observed performance.

\begin{algorithm}[t]
\caption{Evaluation Pipeline in \ourmodel}
\label{alg:protdbench}
\small
\begin{algorithmic}
\REQUIRE Target $(T_{\mathrm{seq}},T_{\mathrm{str}})$; generated binder backbones $\mathcal{D}_{\mathrm{bb}}=\{B_{\mathrm{str}}^{(i)}\}_{i=1}^{N}$;
; verifier $\mathcal{V}$; filter $F(\cdot)$
\ENSURE Metrics $\mathcal{M}$

\STATE \textit{// Sequence evaluation}
\FOR{each backbone $B_{\mathrm{str}}^{(i)} \in \mathcal{D}_{\mathrm{bb}}$}
  \STATE Sample $m$ sequences $\mathcal{S}_i$; evaluate each $s\in\mathcal{S}_i$ with $\mathcal{V}$ to obtain pass indicators $F(s)$
\ENDFOR

\STATE \textit{// Filtering}
\STATE Compute per-sequence success rate from $\{F(s)\}$ over all sampled sequences
\STATE Define the passing backbone set $\mathcal{D}_{\mathrm{pass}}=\{B_{\mathrm{str}}^{(i)} \mid \exists\, s\in\mathcal{S}_i: F(s)=1\}$

\STATE \textit{// Diversity and aggregation}
\STATE Cluster only $\mathcal{D}_{\mathrm{pass}}$ by structural similarity (FoldSeek/TM-score)
\STATE Aggregate $\mathcal{M}$ (sequence-level SR, cluster-based diversity, throughput-aware yield)

\end{algorithmic}
\end{algorithm}

\subsection{Problem Formulation and Evaluation Scope}

We consider the task of \textit{protein binder design} under a unified generative and evaluation setting.
Given a target protein $\{T_{\mathrm{seq}}, T_{\mathrm{str}}\}$, the objective is to generate a binder sequence $B_{\mathrm{seq}}$ together with a corresponding three-dimensional structure $B_{\mathrm{str}}$, such that the binder forms a stable and specific complex with the target under binding constraints $c$ (e.g., predefined interaction hotspots).
Existing generative approaches aim to learn a conditional mapping
\[
f_{\theta}(T_{\mathrm{seq}}, T_{\mathrm{str}}, c) \rightarrow (B_{\mathrm{seq}}, B_{\mathrm{str}}).
\]

For a fixed target, a generative method produces a set of candidate binders
\[
\mathcal{D}_\text{seq} = \{(B_{\mathrm{seq}}^{(i)}, B_{\mathrm{str}}^{(i)})\}_{i=1}^{N},
\]
where each element corresponds to one generated binder sequence–structure pair.
The role of evaluation is to map the generated set $\mathcal{D}_\text{seq}$ to quantitative measures that reflect design quality under a specified evaluation protocol.

To enable systematic and comparable evaluation, we characterize generated binders through a set of complementary properties.
First, \textbf{\emph{per-sequence success rate (SR)}} measures the fraction of
generated sequences that pass a fixed structure-based verification and
filtering protocol, serving as a proxy for both structural validity and
interface quality.
Second, \textbf{\emph{cluster-level pass rate}} quantifies diversity via a cluster-based metric computed from successful binder backbones, capturing
structural diversity beyond redundant solutions.
Finally, \textbf{\emph{structural consistency}} assesses whether a generated
backbone can be recapitulated by an independent structure predictor after
sequence design, reflecting the robustness of generated structures.

\subsection{Evaluation Verifier}

A central component of \ourmodel is the \emph{evaluation verifier}.
An evaluation verifier assigns confidence scores to a binder–target pair, typically based on structure prediction.
Formally, a verifier $\mathcal{V}$ maps
\[
\mathcal{V}\bigl(T_{\mathrm{seq}}, T_{\mathrm{str}}, B_{\mathrm{seq}}^{(i)}, B_{\mathrm{str}}^{(i)}\bigr) \rightarrow \mathbf{z}^{(i)},
\]
where $\mathbf{z}^{(i)}$ denotes a vector of scalar confidence metrics (e.g., pLDDT, ipTM, ipAE).

In practice, evaluation verifiers are instantiated using structure prediction models and their associated confidence outputs.
\ourmodel treats verifiers as modular and replaceable components: different verifiers, or combinations thereof, can be plugged into the same evaluation protocol without modifying the remaining evaluation steps.
This design enables systematic analysis of verifier bias, complementarity, and computational trade-offs.

\subsection{Filtering and Success Criteria}
\begin{figure*}[t!]
    \centering

    \begin{minipage}[t]{0.96\textwidth}
        \raisebox{8pt}{\textbf{(a)}}\\[-1.5ex]
        \includegraphics[width=\linewidth]{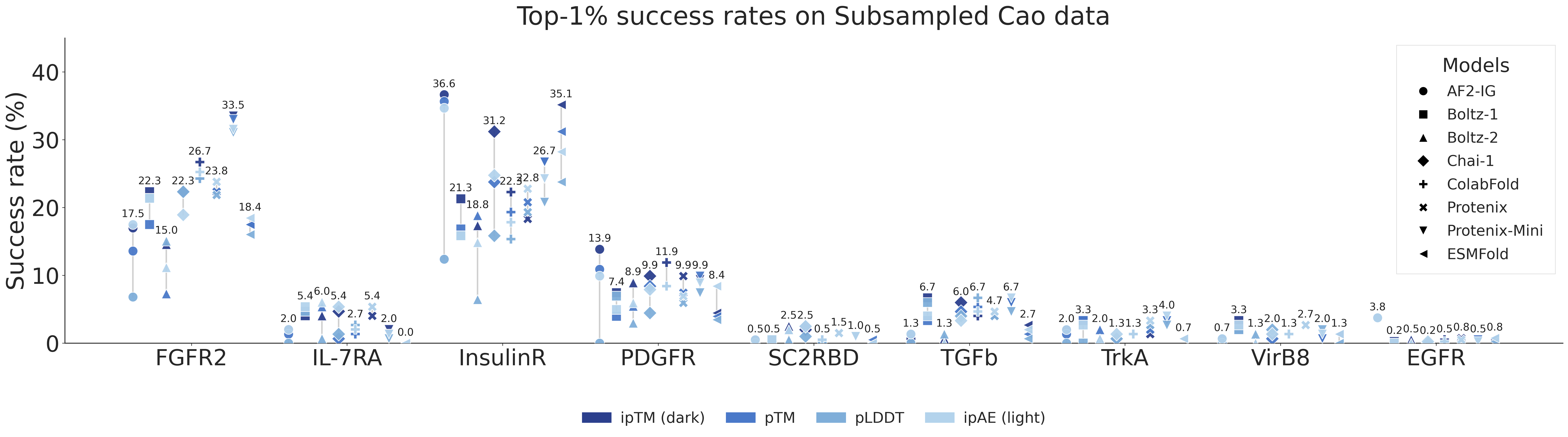}
    \end{minipage}

    \vskip 1.2em

    \noindent
    \begin{minipage}[t]{0.44\textwidth}
        \raisebox{8pt}{\textbf{(b)}}\\[-1.5ex]
        \includegraphics[width=\linewidth]{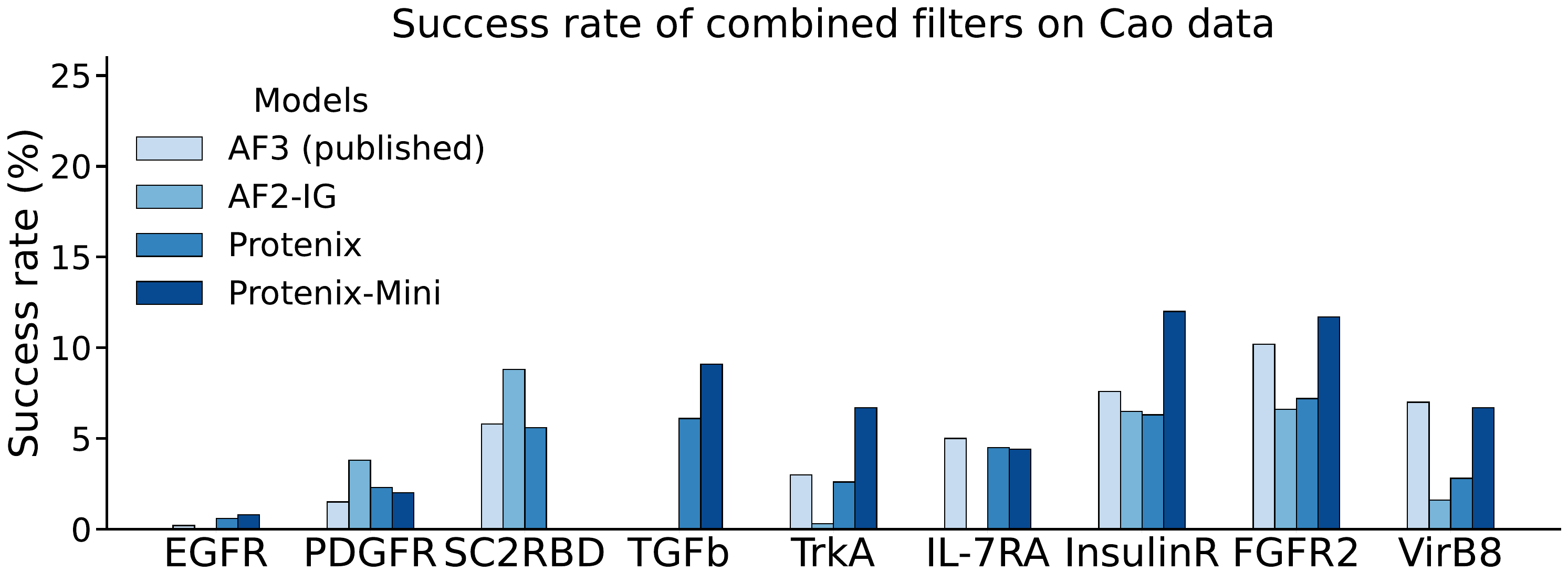}
    \end{minipage}
    \hfill
    \begin{minipage}[t]{0.52\textwidth}
        \raisebox{8pt}{\textbf{(c)}}\\[-1.5ex]
        \includegraphics[width=\linewidth]{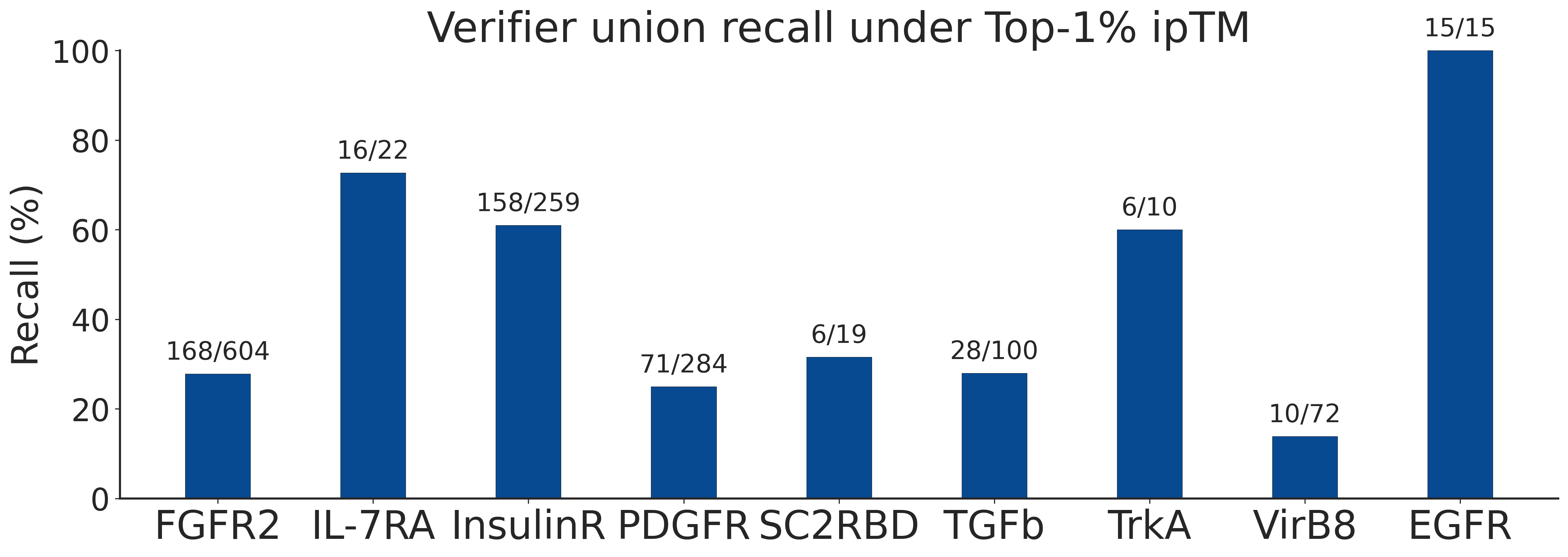}
    \end{minipage}

    \vskip 1.2em

    \begin{minipage}[t]{0.96\textwidth}
        \raisebox{8pt}{\textbf{(d)}}\\[-1.5ex]
        \includegraphics[width=\linewidth]{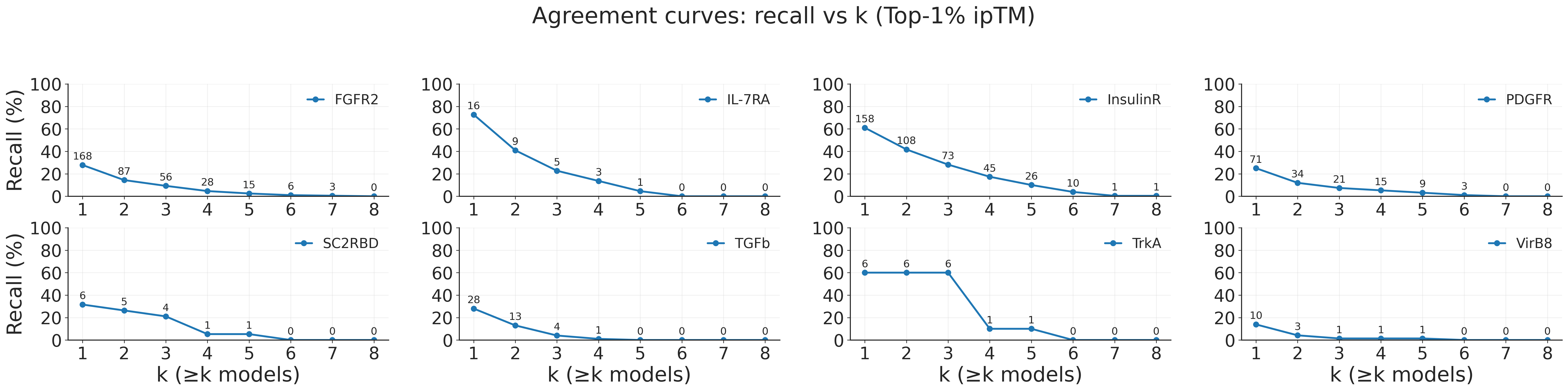}
    \end{minipage}

    \caption{
    \textbf{Benchmarking structure prediction models as filters on the Cao dataset.}
    \textbf{(a)} Top-1\% success rates achieved by individual confidence metrics (e.g., ipTM, pTM, pLDDT, ipAE) derived from AF2-IG, Boltz-1, Boltz-2, Chai-1, ColabFold, Protenix, and Protenix-Mini, on the subsampled Cao set (at most $20{,}000$ negatives per crop).
    \textbf{(b)} Success rates of combined filtering strategies across nine targets (FGFR2, EGFR, IL-7RA, InsulinR, PDGFR, SC2RBD, TGF$\beta$, TrkA, VirB8), evaluated against the full Cao screening pool. ``AF3 (published)'' denotes baseline results from prior work, while other bars correspond to results obtained using our unified evaluation pipeline.
    \textbf{(c)} Union recall under Top-1\% ipTM filtering, defined as the fraction of true positive binders recovered by at least one model, establishing an upper bound on multi-model filtering. Per-target true-positive counts are annotated above each bar (true positives in union / total wet-lab binders).
    \textbf{(d)} Per-target agreement curves under Top-1\% ipTM filtering: recall (\%) as a function of the agreement threshold $k$, where a true positive must be identified by at least $k$ verifiers simultaneously. Recall decreases rapidly as $k$ increases, indicating limited consensus among verifiers and highlighting their complementary coverage of the binder design space. The full nine-target version is provided in \Cref{fig:agreement-curves}.
    }
    \label{fig:filter-benchmark}
\end{figure*}

Within \ourmodel, evaluation is defined as a deterministic mapping from a set of generated binders to a set of quantitative performance measures.
This mapping is specified through two conceptually distinct components: \emph{filtering} and \emph{success criteria}.

\paragraph{Filtering}
Given the verifier $\mathcal{V}$ introduced above, which assigns confidence scores to a target-binder pair
\[
\mathcal{V}(T_{\mathrm{seq}}, T_{\mathrm{str}}, B_{\mathrm{seq}}^{(i)}, B_{\mathrm{str}}^{(i)}) \rightarrow \mathbf{z}^{(i)} \in \mathbb{R}^k.
\]
A \emph{filter} is defined as a deterministic predicate
$
\mathcal{F}(\mathbf{z}) \in \{0,1\},
$
which maps verifier scores to a binary decision.
In practice, $\mathcal{F}$ is instantiated using threshold-based rules on verifier scores.
For example, AlphaProteo~\citep{zambaldi2024novo} applies a conjunction of interface confidence and structural consistency thresholds, e.g., $\mathrm{pTM}_\mathrm{binder} > 0.8$, $\mathrm{RMSD} < 2.5$ and $\min(\mathrm{ipAE}) < 1.5$.

\paragraph{Success criteria}
Given a generated set of binder backbones $\mathcal{D}$, we define success criteria to summarize performance into metrics after applying a verifier-based filter.

At the \emph{binder sequence level}, we report the per-sequence pass rate as the fraction of sampled sequences that pass the filter. At the \emph{binder cluster level}, we quantify diversity among successful designed backbones by clustering passing backbones based on structural similarity (pairwise TM-score; FoldSeek~\citep{van2024fast}) and report a cluster-based success metric described in \Cref{binder_diversity}.

\subsection{Throughput-Aware Evaluation}

Beyond success-rate oriented metrics, \ourmodel incorporates computational throughput as an additional evaluation dimension.
Given a fixed computational budget, e.g., $24$-hour single GPU computation,
performance metrics are measured for 
fair comparison among methods with different computational profiles. Throughput-aware evaluation reflects practical deployment constraints in large-scale binder design processes, where both design quality and computational cost jointly determine real-world utility.




\section{Analyzing Structure Prediction Models as Evaluation verifiers}
\label{sec:filter}
Under the evaluation framework specified in Section
~\ref{sec:framework}, structure prediction models act as \emph{evaluation verifiers} that determine which generated candidates are counted as successful binders.
The choice of verifier directly affects evaluation outcomes by shaping binder enrichment, computational cost, and coverage of the design space.
As such, commonly-used structure predictors are analyzed as evaluation verifiers using wet-lab data, focusing on filtering accuracy, efficiency, and design space coverage.

\begin{table*}[t] 
\centering
\caption{\textbf{Thresholds for combined filters.}
``AF2-IG-easy'' reflects thresholds proposed by BindCraft~\citep{pacesa2025one}.
``AF2-IG'' denotes thresholds selected via our own grid search on Cao data; these match the values independently reported in~\citet{zambaldi2024novo}.
``AF3'' refers to thresholds grid-searched in~\citet{zambaldi2024novo}.
``Protenix'', ``Protenix-Mini'' share a unified threshold set, and thresholds are also selected via grid search on Cao data.
The three structural thresholds are backbone RMSDs computed differently:
\textbf{complex RMSD} is taken over the full binder--target complex backbone, comparing the designed complex against the verifier's prediction after sequence design;
\textbf{binder RMSD} is taken over the binder backbone only, comparing the generated binder backbone against the verifier's prediction after sequence design;
\textbf{binder bound/unbound RMSD} is the backbone RMSD between the binder predicted in isolation and the same binder predicted jointly with the target.}
\label{tab:filter-thresholds}

\scalebox{0.8}{
\begin{tabular}{l|ll}
\toprule
\textbf{Filter Name} & \textbf{Confidence Thresholds} & \textbf{Structure Thresholds} \\
\midrule
AF3~\citep{zambaldi2024novo} & min ipAE $<$ 1.5, binder pTM $>$ 0.8 & complex RMSD $<$ 2.5 Å \\
AF2-IG~\citep{zambaldi2024novo} & ipAE $<$ 7.0, pLDDT $>$ 0.9 & binder RMSD $<$ 1.5 Å \\
AF2-IG-easy~\citep{pacesa2025one} & ipAE $<$ 10.85, ipTM $>$ 0.5, pLDDT $>$ 0.8 & binder bound/unbound RMSD $<$ 3.5 Å \\
\midrule
\makecell[l]{Protenix~\citep{chen2025protenix},\\Protenix-Mini~\citep{gong2025protenix}} & binder ipTM $>$ 0.85, binder pTM $>$ 0.88 & complex RMSD $<$ 2.5 Å \\
\bottomrule
\end{tabular}
}
\vspace{-10pt}
\end{table*}

\subsection{Analysis Setup}

Filtering is a central operation in practical binder design pipelines.
To serve as reliable evaluation verifiers, structure-predictor's confidence metrics are expected to have correlations with wet-lab experimental binding success while remaining computationally feasible at scale.

\paragraph{Dataset}
We use the Cao dataset~\citep{cao2022design} to build the evaluation environment.
Cao dataset is a large wet-lab annotated benchmark consisting of a number of energy-based model designed \emph{de novo} binder candidates for multiple targets.
Each candidate is experimentally labeled as a \emph{binder} or \emph{non-binder}.
Crucially, these designs were validated without prior AlphaFold-based filtering, making the dataset a suitable standard for retrospectively assessing the correlation between \emph{in silico} metrics and \emph{in vitro} results.

\paragraph{Evaluation verifiers and filters}
We evaluate confidence-based filtering strategies derived from multiple families of structure predictors.
Specifically, we consider:
\begin{itemize}
    \item \textbf{AF2-style models}, represented by AF2-IG~\citep{watson2023novo}, 
    a variant of AF2 adapted for binder design using an initial-guess protocol, 
    and ColabFold~\citep{mirdita2022colabfold}, an accelerated AF2 inference pipeline 
    that replaces the original MSA search with MMseqs2.
    \item \textbf{AF3-style models}, represented by Protenix~\citep{chen2025protenix}, its lightweight variant Protenix-Mini~\citep{gong2025protenix}\footnote{Protenix-Mini employs a two-step ODE sampler for efficient inference.}, Boltz-1~\citep{wohlwend2024boltz}, Boltz-2~\citep{passaro2025boltz2}, Chai-1~\citep{chai2024chai}.
    \item \textbf{MSA-free model}, ESMFold~\citep{lin2023evolutionary}.
\end{itemize}

From each verifier, we extract commonly used confidence metrics, including ipTM, ipAE, pLDDT, and pTM.
For each verifier and target, a single-metric filter is constructed by retaining the top $\alpha$ fraction of candidates under a confidence score $s(B)$, with $\alpha=1\%$.
Formally, for a fixed target and a given confidence metric $s(B)$ produced by a verifier,
candidates are filtered according to $s(B)$.
A binary filter is defined by retaining the top $\alpha$ fraction:
\begin{equation}
\label{eq:top1-filter}
\mathcal{F}_{\alpha}(B)
=
\mathbb{I}
\left[
s(B)
\ge
\mathrm{Quantile}_{1-\alpha}
\bigl(\{s(B')| B' \in \mathcal{D} \}\bigr)
\right].
\end{equation}
This provides a simple, model-agnostic baseline for assessing verifier enrichment behavior. 
In practice, for computational tractability, non-binders are randomly subsampled to at most $20000$ per target.

Other than filters induced by single confidence metric, combined filters apply multiple score thresholds jointly (e.g., interface confidence and structural consistency constraints) to classify candidates as passing or failing.
Thresholds are selected via a simple grid search on the Cao dataset, following prior work~\citep{zambaldi2024novo}. Based on the resulting single-metric analysis, we identify Protenix as a strong and representative evaluation verifier for combined filter studies. The resulting fixed configurations are summarized in Table~\ref{tab:filter-thresholds}, with full details provided in Appendix~\ref{appendix:filter-search}.

\paragraph{Filtering accuracy metric.}
Given experimental binding labels $y(B)\in\{0,1\}$, where $y(B)=1$ indicates that candidate $B$ was experimentally confirmed as a binder by~\citet{cao2022design},
we quantify filtering accuracy as the \emph{success rate of retained candidates}: the fraction of designs retained by the filter $\mathcal{F}_{\alpha}$ that are experimentally validated binders.
Concretely, the numerator counts retained designs that are wet-lab binders, and the denominator counts all retained designs:
\begin{equation}
\label{eq:filter-acc}
\mathrm{SR}
=
\frac{
\sum_{B \in \mathcal{D}}
\mathcal{F}_{\alpha}(B)\, y(B)
}{
\sum_{B \in \mathcal{D}} \mathcal{F}_{\alpha}(B)
}.
\end{equation}
A higher $\mathrm{SR}$ thus means the verifier's top-$\alpha$ ranking is more enriched for true binders.

\paragraph{Evaluation questions.}
Using this setup, we analyze evaluation verifiers along three dimensions:
\begin{enumerate}[left=5pt]
    \item \emph{\textbf{Filtering accuracy}: How effectively does a verifier enrich experimentally validated binders?}
    \item \emph{\textbf{Efficiency}: Can lightweight verifiers reduce computational cost without sacrificing enrichment quality?}
    \item \emph{\textbf{Coverage}: Do different verifiers capture complementary regions of the design space?}
\end{enumerate}

\subsection{Main Findings}

\textbf{Filtering accuracy: verifier choice substantially affects binder enrichment.}
We separate filtering accuracy into two complementary questions, reported respectively in Figure~\ref{fig:filter-benchmark}(a) and (b).
Panel (a) measures \emph{single-score ranking ability} on the subsampled Cao set (at most $20{,}000$ negatives per crop) at the Top-1\% threshold, while panel (b) measures \emph{practical combined-filter screening ability} on the full Cao pool.
Because the full set contains substantially more negatives, absolute success rates in (b) are systematically lower than in (a); this reflects realistic class imbalance rather than a contradiction between the panels, and the two panels are therefore complementary rather than directly comparable.

In Figure~\ref{fig:filter-benchmark}a we observe substantial variability across both models and targets.
AF3-style predictors, especially Protenix and Protenix-Mini, achieve higher top-1\% success rates than AF2-IG and perform competitively compared to other folding models on most targets, indicating strong single-metric enrichment capability.
However, no single metric or verifier dominates uniformly across all targets.
In Figure~\ref{fig:filter-benchmark}b, Protenix-based combined filters remain competitive compared to AF3 and AF2-IG, while the lightweight Protenix-Mini retains comparable or even superior performance to Protenix.

\textbf{Efficiency: lightweight verifier variants enable scalable evaluation.}
Protenix-Mini substantially reduce inference time relative to the full Protenix while maintaining comparable enrichment performance (Figure~\ref{fig:filter-benchmark}a).
This cost accuracy trade-off highlights that verifier efficiency is a critical factor in large-scale evaluation, where computational budget constraints can otherwise dominate practical feasibility.

\textbf{Coverage: evaluation verifiers emphasize complementary regions of the design space.}
Despite exhibiting broadly similar enrichment trends, 
different evaluation verifiers recover largely distinct subsets of true binders under identical filtering criteria.
As shown in Figure~\ref{fig:filter-benchmark}c, 
taking the union of Top-1\% ipTM predictions across verifiers substantially improves recall for several targets.
It indicates that no single verifier dominates coverage of the true binder space.
Moreover,
as the agreement threshold increases,
recall drops sharply (Figure~\ref{fig:filter-benchmark}d).
This fact demonstrates that only a small fraction of true binders are consistently identified by multiple verifiers.
Together, 
these results indicate that different structure prediction models potentially capture complementary signals rather than redundant confidence information.
We suggest that it may reflect distinct inductive biases toward specific structural features or interaction patterns in the binder design space.
Consequently, evaluation outcomes can vary substantially depending on the choice of verifier.

\section{Evaluating Generative Models within the \ourmodel Framework}
\label{sec:gen_benchmark}

Having established the evaluation framework in  Section~\ref{sec:framework} and analyzed the behavior of structure prediction models as evaluation verifiers in Section~\ref{sec:filter}, we now turn to benchmark different generative models performance under a fixed protocol.
This section presents a systematic evaluation of generative binder design models under a fixed protocol, enabling controlled comparison of generation efficiency, per-sequence success, structural diversity, and structural consistency. Rather than establishing a definitive ranking of generative models, the analysis illustrates how different evaluation choices within \ourmodel expose distinct trade-offs between throughput, success rate, and diversity.

\subsection{Benchmark Setup}

\paragraph{Benchmark targets}
\label{benchmark_settings}
Following prior work~\citep{zambaldi2024novo}, 
we benchmark on 10 protein targets with diverse structural properties and varying design difficulty.
The target set includes multi-chain complexes (H1, VEGF-A, IL17A, and TNFa), 
which further challenge generative models with complex interfaces and cooperative binding geometries.
For each target, 
we follow the same cropping ranges, hotspot specifications, and binder length settings across all methods.
Detailed target information is provided in Appendix~\ref{target_infomation}.
For each target and binder length, 
we generate 4 independent backbone structures per method to reduce variance and ensure consistent comparison.

\paragraph{Generative binder design models}
Several recent binder design methods, such as Latent-X~\citep{bridgland2025latent}, AlphaProteo~\citep{zambaldi2024novo}, Chai-2~\citep{chai2025zero}, and SeedProteo~\citep{qu2025seedproteoaccuratenovoallatom}, do not release code or pretrained weights and are therefore excluded from head-to-head benchmarking.
We evaluate representative open-source methods that support hotspot-conditioned binder design:
\begin{itemize}[left=5pt]
    \item \textbf{Diffusion-based models}: RFdiffusion-3~\citep{butcher2025novo}, BoltzGen~\citep{Stark2025BoltzGen}, Protpardelle-1~\citep{chu2023allatom}, ODesign~\citep{zhang2025odesignworldmodelbiomolecular}, and PXDesign~\citep{ren2025pxdesign}.
    \item \textbf{Hallucination-based models}: BindCraft~\citep{pacesa2025one} and BoltzDesign1~\citep{cho2025boltzdesign1}.
\end{itemize}
All methods are provided with the same target sequences, structures and identical hotspot constraints.

\begin{table*}[t]
\centering
\caption{
The number of \textbf{successful backbone structures} produced per 24 hours on a single NVIDIA A100 GPU.
All results are reported under a unified end-to-end budget that includes the full pipeline (e.g., generation, inverse folding, and evaluation).
A backbone structure is counted as successful if it yields $\geq 1$ sequence passing the AF2-IG-Easy filter. 
}

\label{sample-table-af2-comprehensive}
\begin{small}
\setlength{\tabcolsep}{4pt}
\scalebox{0.8}
{
\begin{tabular}{ll|cccccccccc}
\toprule
\multicolumn{2}{c|}{\textbf{Models}} &
\textbf{BHRF1} & \textbf{PDL1} & \textbf{IR} & \textbf{TrkA} & \textbf{IL7RA} &
\textbf{SC2RBD} & \textbf{VEGFA} & \textbf{H1} & \textbf{IL17A} & \textbf{TNFa} \\
\midrule
\multirow{5}{*}{\textbf{Diffusion}}
& \textbf{RFDiffusion-3} & 751 & 542 & 512 & 407 & 182 & 131 & 98 & 30 & \textbf{37} & 0 \\
& \textbf{Boltzgen}    & 428 & 549 & 606 & \textbf{804} & 279 & 34 & 207 & \textbf{248} & 15 & 15 \\
& \textbf{Protpardelle-1c}     & 142 & 403 & 18 & 248 & 9 & 35 & 3 & 11 & 0 & 0 \\
& \textbf{ODesign}     & \textbf{901} & 375 & 559 & 480 & 224 & 127 & 152 & 115 & 3 & 0 \\
& \textbf{PXDesign}    & 837 & \textbf{1247} & \textbf{756} & 758 & \textbf{695} & \textbf{295} & \textbf{464} & 193 & 31 & \textbf{57} \\
\midrule
\multirow{2}{*}{\textbf{Hallucination}}
& \textbf{BoltzDesign-1} & 42 & 54 & 51 & 73 & 17 & 15 & 13 & 4 & 5 & 0 \\
& \textbf{BindCraft}     & 122 & 231 & 220 & 194 & 89 & 70 & 41 & 28 & 28 & 1 \\
\bottomrule
\end{tabular}
}
\end{small}
\end{table*}

\begin{figure*}[t]
    \centering
    \begin{minipage}[t]{0.48\linewidth}
        \raisebox{8pt}{\textbf{(a)}}\\[-1.5ex]
        \includegraphics[width=\linewidth]{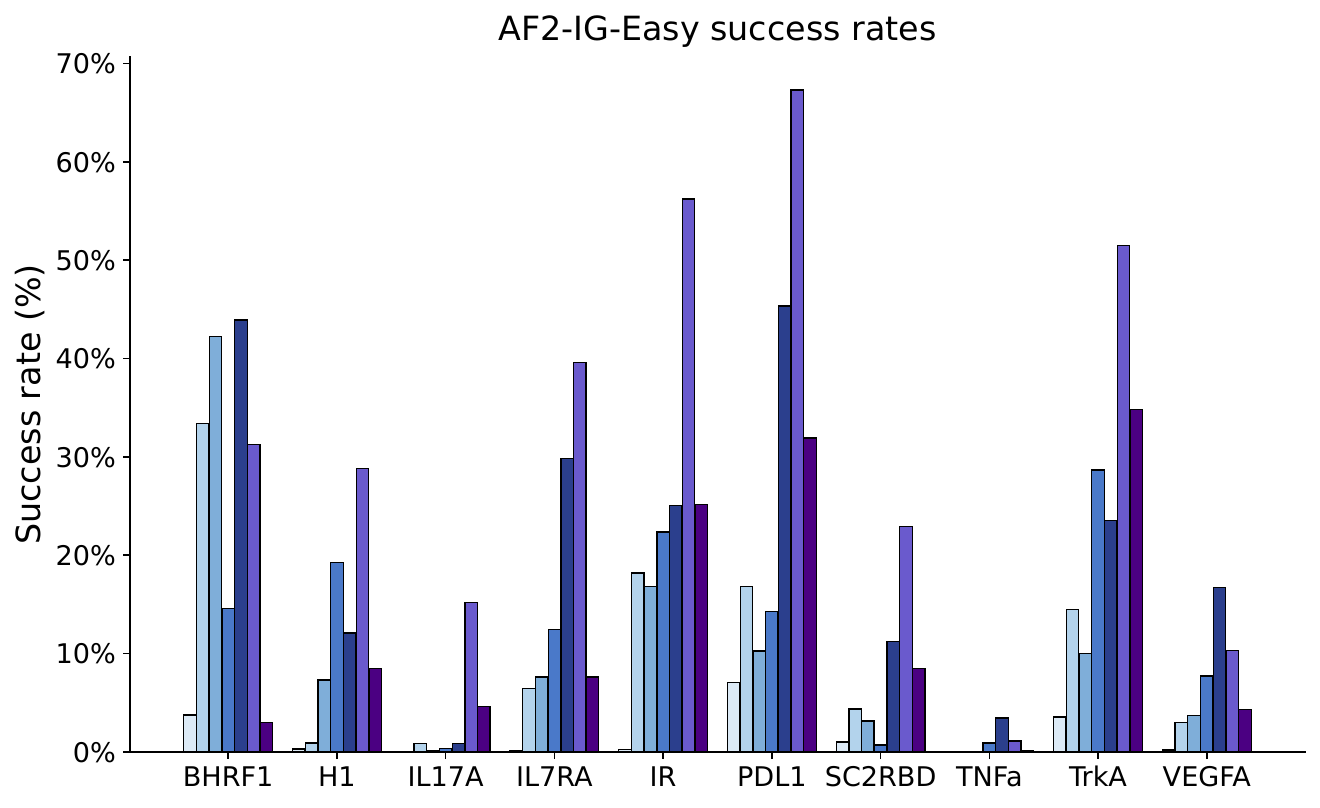}
    \end{minipage}
    \hfill
    \begin{minipage}[t]{0.48\linewidth}
        \raisebox{8pt}{\textbf{(b)}}\\[-1.5ex]
        \includegraphics[width=\linewidth]{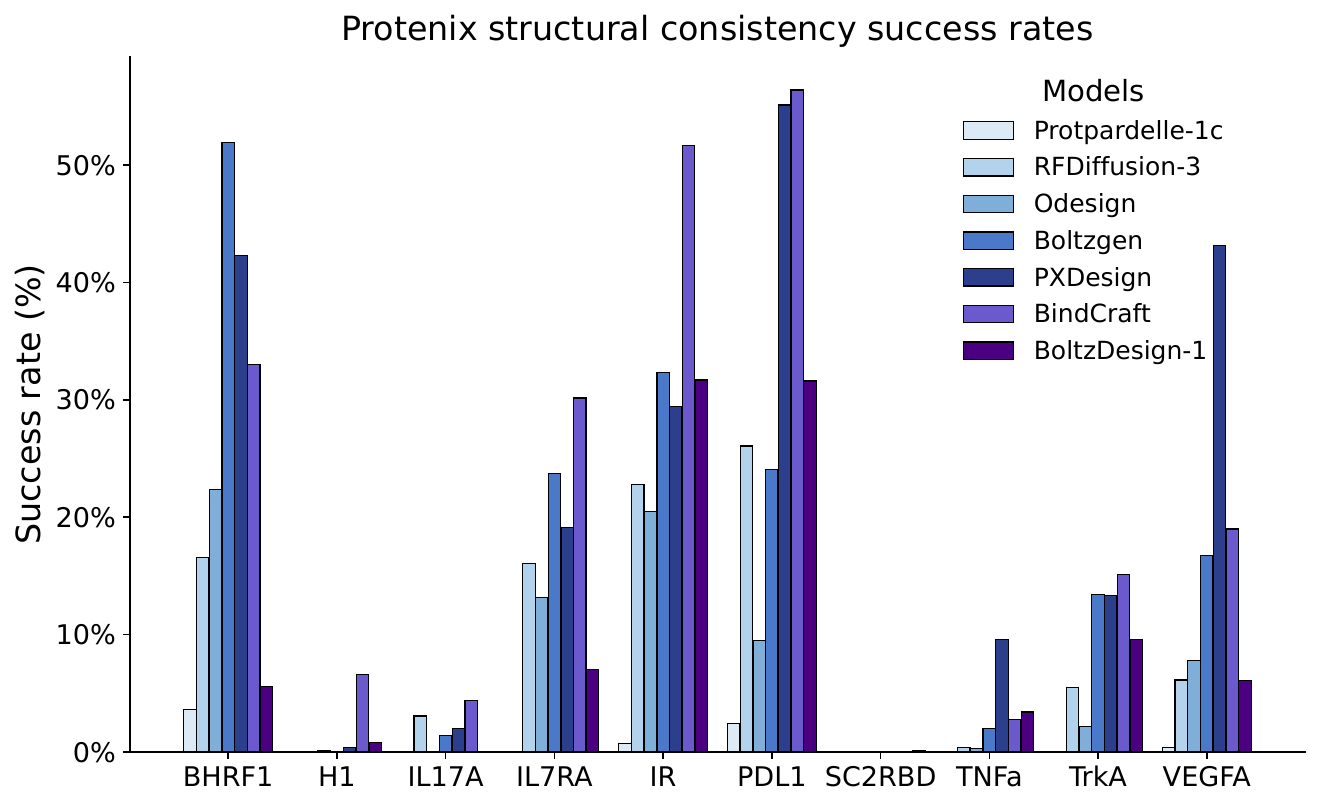}
    \end{minipage}
    \caption{
    \textbf{Per-sequence binder design benchmarks across targets.}
    \textbf{(a)} Per-sequence success rate (\%) under the AF2-IG-Easy filter (\Cref{tab:filter-thresholds}).
    \textbf{(b)} Per-sequence structural consistency (\%), measured as Protenix-Mini recapitulation of the designed backbone after sequence design.
    }
    \label{fig:combined-large}
\end{figure*}

\begin{figure*}[t]
    \centering
    \vspace{-5pt}
    \includegraphics[width=\linewidth]{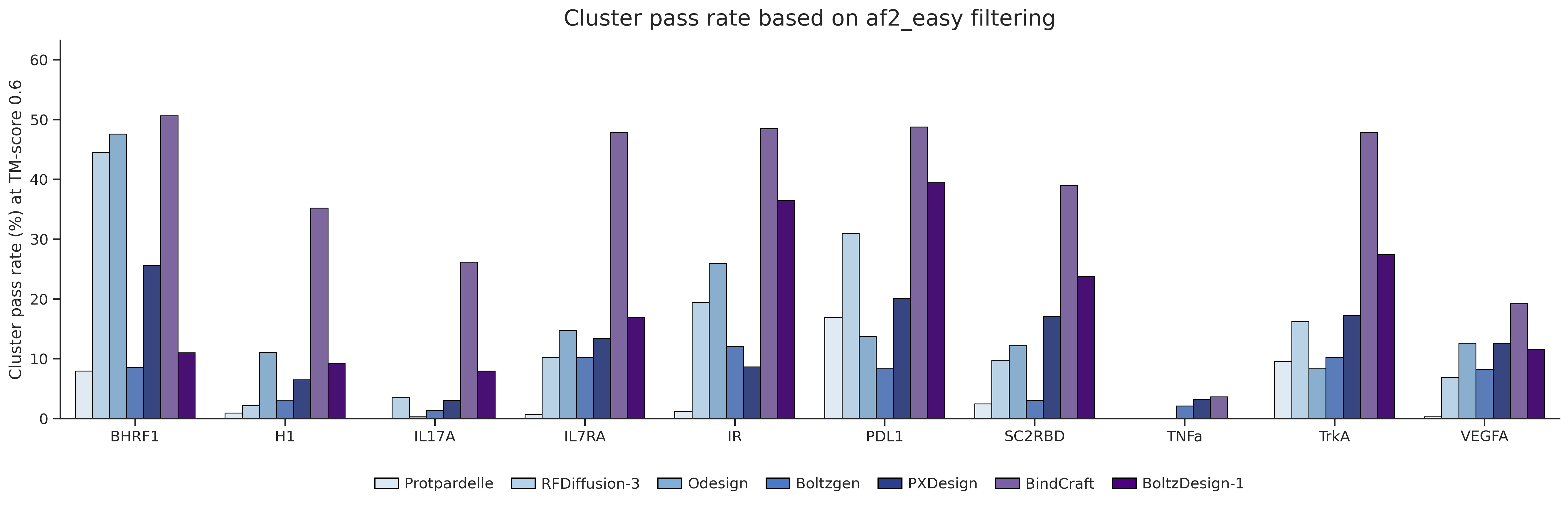}
    \vspace{-15pt}
    \caption{
    \textbf{Binder structural diversity benchmark (TM-score $\geq 0.6$).}
    For each target, bars show the \emph{diversity-adjusted cluster pass rate} of each method after AF2-IG-Easy filtering: the number of unique structural clusters at TM-score threshold $0.6$ among passing backbones, normalized by the total number of designed backbones. The complete results across TM-score thresholds $0.6$, $0.8$, and $1.0$ are reported in \Cref{app:diversity-other-tm}.
    }
    \label{fig:binder_diversity}
    \vspace{-5pt}
\end{figure*}
\subsection{Evaluation protocol and metrics}
Different generative paradigms trade off sampling speed and per-sample success rate.
To reflect practical design workflows, \ourmodel reports both \emph{generation throughput} and \emph{per-sequence success} explicitly. For generative model benchmarking, we choose the widely used AF2-IG-Easy as evaluation verifier to ensure fair and comparable assessment across methods\footnote{Results with protenix-mini as evaluation verifier are provided in \Cref{app:protenix-mini-benchmark}}. 

\textbf{Generation throughput (successful structures per 24 hours)}
We measure generation efficiency under a fixed 24-hour budget on a single NVIDIA A100 GPU.
This budget includes backbone generation time, downstream inverse folding, and structure-based verifier evaluation.
A generated backbone structure is counted as successful if it yields at least one sequence that passes the AF2-IG-Easy filter
(see Table~\ref{tab:filter-thresholds} for details).
This metric captures how many structurally viable binder candidates a method can deliver under realistic resource constraints.
We measure the total GPU time on each target as
$
T_{\mathrm{tot}} = T_{\mathrm{sample}} + T_{\mathrm{MPNN}} + T_{\mathrm{eval}},
$
where the three terms are the total time spent on backbone generation \footnote{We note that BindCraft integrates generation and internal scoring; therefore we run BindCraft without internal scoring (see \Cref{appendix:implementation_details}), and results should be interpreted as a throughput proxy rather than strictly identical end-to-end runtime across pipelines.}, inverse folding (over all sampled sequences), and verifier evaluation (over all sampled sequences), respectively.
We then estimate the expected number of successful backbones produced in 24 hours as
\begin{equation}
\label{eq:n24h}
N_{24h}
=
\frac{24\ \mathrm{hours}}{T_{\mathrm{tot}}}
\cdot
N_{\mathrm{pass}},
\end{equation}
$N_{\mathrm{pass}}$ is the number of passing backbones (a backbone passes if any of its associated sequences passes the filter).

\textbf{Per-sequence success rate (SR)}
For each generated backbone structure $B_{\mathrm{str}}^{(i)}$, we sample $m=8$ sequences
$\mathcal{S}_i$ using ProteinMPNN.
Each sequence is evaluated independently with the AF2-IG-Easy protocol.
A sequence is considered successful if its score exceeds the filter threshold $\tau$,
\begin{equation}
\label{eq:seq-filter}
\mathcal{F}(s) = \mathbb{I}\bigl[\mathrm{Score}(s, B_{\mathrm{str}}) \ge \tau\bigr].
\end{equation}

We define per-sequence success rate (SR) as the overall pass rate across all sampled sequences,
\begin{equation}
\label{eq:sr-micro}
\mathrm{SR}
=
\frac{
\sum_{i=1}^{N} \sum_{s \in \mathcal{S}_i} \mathcal{F}(s)
}{
\sum_{i=1}^{N} |\mathcal{S}_i|
}.
\end{equation}

\textbf{Diversity of successful binders}
\label{binder_diversity}
To quantify the structural diversity of successful binders, we cluster only the backbones that pass the AF2-IG-Easy filter using Foldseek.
For each generated backbone $B_{\mathrm{str}}^{(i)}$ with sampled sequences $\mathcal{S}_i$, we define backbone success as
$\mathcal{F}_{\mathrm{bb}}(i)=\mathbb{I}\bigl[\exists\, s\in\mathcal{S}_i:\mathcal{F}(s)=1\bigr]$ (i.e., any of $m{=}8$ sequences passes).
Let $\mathcal{D}_{\mathrm{pass}}=\{B_{\mathrm{str}}^{(i)}\mid \mathcal{F}_{\mathrm{bb}}(i)=1\}$.
Foldseek clustering on $\mathcal{D}_{\mathrm{pass}}$ yields structural clusters $\mathcal{C}_{\mathrm{pass}}$, and we define
\begin{equation}
\label{eq:cluster-success}
\mathrm{S}_{\mathrm{cluster}} = \left| \mathcal{C}_{\mathrm{pass}} \right|.
\end{equation}
We further report a diversity-adjusted success rate
$\mathrm{R}_{\mathrm{cluster}} = \mathrm{S}_{\mathrm{cluster}} / N_{\mathrm{total}} \times 100\%$,
where $N_{\mathrm{total}}$ is the total number of generated backbones.

\textbf{Structural consistency}
As an auxiliary evaluation signal, we assess whether generated backbone structures can be recapitulated by an independent structure predictor after sequence design.
Formally, for a designed sequence $s$ from backbone $B_{\mathrm{str}}^{_{\mathrm{backbone}}}$,
let $\hat{B}_{\mathrm{str}}^{_{\mathrm{backbone}}}(s)$ denote the structure predicted by Protenix-Mini. 
We define a structural consistency indicator as,
\begin{equation}
\label{eq:struct-consistency}
\mathcal{C}(s)
=
\mathbb{I}
\Bigl[
\operatorname{RMSD}\!\bigl(
B_{\mathrm{str}}^{\mathrm{backbone}},
\hat{B}_{\mathrm{str}}^{\mathrm{backbone}}(s)
\bigr)
< \delta
\Bigr].
\end{equation}

with $\delta = 2.5$\,\AA.
We report structural consistency rate (CR),
defined as the fraction of sequences whose predicted structures
recapitulate the generated backbone:
\begin{equation}
\label{eq:cr}
\mathrm{CR}
=
\frac{1}{\sum_i |\mathcal{S}_i|}
\sum_{i} \sum_{s \in \mathcal{S}_i} \mathcal{C}(s).
\end{equation}

\subsection{Main Findings}
\textbf{Generation throughput varies substantially across generative paradigms.}
Table~\ref{sample-table-af2-comprehensive} reports the number of successful backbone structures under a standardized evaluation protocol. In our benchmark, 
all methods are evaluated using a comparable number of generated backbones. 
We apply the same downstream inverse folding and filtering pipeline.
This pipeline allows us to isolate differences in generation quality and evaluation behavior. 
Under this controlled setting, 
diffusion-based methods consistently achieve orders-of-magnitude higher throughput than hallucination-based approaches across most targets. Among them, 
BoltzGen and PXDesign produce the largest number of successful structures on most of the targets, 
while extremely challenging cases such as TNFa remain difficult for several methods. 
We note that this protocol is designed for controlled comparison rather than optimal deployment.
In practical production settings, 
diffusion-based and hallucination-based methods would potentially adopt different allocations of computational budget between backbone generation and evaluation.
Exploring such strategy-dependent budget allocation is an important direction for future large-scale design pipelines, 
but is beyond the scope of the present benchmark.

\textbf{Throughput and per-sequence success rate are decoupled.}
Per-sequence success rate (Figure~\ref{fig:combined-large}a) measures \emph{sampling quality}: conditional on a generated backbone, how often ProteinMPNN+AF2-IG-Easy returns a passing sequence.
Throughput (Table~\ref{sample-table-af2-comprehensive}) measures \emph{compute-amortized yield}: under a fixed 24-hour single-A100 budget, how many distinct successful backbones a method actually produces.
Because $N_{24\mathrm{h}}$ scales as $T_{\mathrm{tot}}^{-1}\cdot N_{\mathrm{pass}}$ (\Cref{eq:n24h}), these two axes can move independently: a high per-sequence success rate does not guarantee high throughput when $T_{\mathrm{tot}}$ is large, and a low per-sequence success rate does not preclude high throughput when sampling is cheap.
RFdiffusion-3 illustrates this: it attains competitive per-sequence success on most targets, but its substantially higher per-sample cost yields lower 24-hour throughput than BoltzGen and PXDesign.
BoltzGen and PXDesign instead strike a more favorable balance, combining strong per-sequence success with substantially higher throughput.
Hallucination-based methods such as BindCraft exhibit high per-sequence success rates;
however, this is partly attributable to their integrated generation-and-filtering pipelines, where candidates failing intermediate quality thresholds (e.g., low ipTM) are iteratively regenerated, biasing the final sample set toward higher post-hoc success.
Reporting only one of these two axes can therefore misrepresent practical method performance, and we report both throughout \ourmodel.

\begin{table}[t]
\centering
\small
\caption{\textbf{End-to-end runtime breakdown by method.}
Percentage of total end-to-end runtime $T_{\mathrm{tot}}$ spent on generation ($T_{\mathrm{sample}}$) vs.\ downstream evaluation ($T_{\mathrm{MPNN}} + T_{\mathrm{eval}}$, with $T_{\mathrm{MPNN}}$ negligible), averaged across the ten benchmark targets.}
\label{tab:runtime-breakdown}
\begin{tabular}{lrr}
\toprule
\textbf{Method} & \textbf{Gen.\ (\%)} & \textbf{Eval.\ (\%)} \\
\midrule
Protpardelle-1c & 2.83  & 97.17 \\
BoltzGen        & 7.54  & 92.46 \\
PXDesign        & 11.56 & 88.44 \\
ODesign         & 13.43 & 86.57 \\
RFDiffusion-3   & 33.88 & 66.12 \\
BindCraft       & 89.04 & 10.96 \\
BoltzDesign-1   & 92.95 & 7.05  \\
\bottomrule
\end{tabular}
\end{table}

\Cref{tab:runtime-breakdown} reports the percentage share of $T_{\mathrm{tot}}$ attributable to generation versus downstream evaluation, averaged over the ten benchmark targets.
The bottleneck is strongly method-dependent.
For diffusion-based methods, downstream evaluation dominates total runtime (Protpardelle, BoltzGen, PXDesign, ODesign all spend $>85\%$ of $T_{\mathrm{tot}}$ on evaluation), reflecting the cost of running ProteinMPNN $\times$ AF2-IG-Easy on every generated backbone.
RFDiffusion-3 sits in between ($66.12\%$ evaluation).
In contrast, hallucination-based methods are generation-dominated (BindCraft $89.04\%$, BoltzDesign-1 $92.95\%$ of $T_{\mathrm{tot}}$ on generation), because each design requires hundreds of inner optimization steps before any sequence is sampled.
This breakdown clarifies why per-sequence success rate and 24-hour throughput can be decoupled: for diffusion methods, throughput is bottlenecked by the verifier rather than the sampler, whereas for hallucination methods the opposite holds.
The measurement protocol is detailed in \Cref{appendix:runtime}.

\textbf{Structural diversity reflects differences in effective design space exploration.}
Figure~\ref{fig:binder_diversity} reports the \emph{diversity-adjusted cluster pass rate} for each method after AF2-IG-Easy filtering, 
computed by clustering only the backbones that pass the filter at multiple TM-score thresholds.
Hallucination-based methods exhibit higher cluster pass rates across targets, 
potentially indicating broader exploration of the structural design space. 
These hallucination-based pipelines initialize each design independently and optimize it without an explicit generative prior.
It leads to exploration of diverse local optima based on structure prediction models' confidence landscape.
In contrast, 
diffusion-based models sample from a learned distribution,
which can lead to more concentrated sampling around a smaller number of structural modes that satisfy the evaluation criteria.

\textbf{Structural consistency highlights robustness of generated scaffolds.}
Figure~\ref{fig:combined-large}b reports structural consistency measured by Protenix-Mini recapitulation.
PXDesign achieves consistently high consistency across diverse targets, indicating that its generated backbones are robust to sequence design.
In contrast, targets such as SC2RBD remain challenging across all methods, suggesting that structural consistency is jointly influenced by generative quality and intrinsic target structure prediction difficulty.

Overall, these results demonstrate that when evaluated under a fixed protocol, generative binder design methods exhibit markedly different trade-offs between efficiency, per-sequence success, and diversity.
\section{Discussion}
In this work, we address a central but underexplored challenge in \emph{de novo} protein binder design: the lack of standardized, and practically meaningful evaluation protocols.

A key finding of our study is that evaluation outcomes are highly sensitive to the choice of structure prediction verifier.
As shown in Section~\ref{sec:filter}, different folding models exhibit distinct inductive biases, leading to markedly different enrichment behavior even under identical filtering protocols.
This cautions against over-interpreting absolute success rates reported under a single verifier and motivates multi-verifier or verifier-aware evaluation in future work.
Our verifier-bias analysis is primarily based on the Cao dataset of energy-based designs; while this enables controlled retrospective evaluation against wet-lab annotations, the observed bias patterns may not fully generalize to binders produced by deep generative models.
A particular concern arises for hallucination-based generation, where the concern runs deeper than a simple model-family overlap.
By construction, these pipelines rely on gradient signals from the loss landscape of a folding model's confidence head, while the optimization itself proceeds in discrete sequence space.
This combination makes it plausible for optimization to collapse into narrow niches of the confidence landscape that score highly under the folding model yet correspond to sequences without genuine biological or biophysical meaning—the discrete-space search can effectively exploit adversarial-like ridges in the verifier's score surface that an unrelated wet-lab assay would not reward.
Evaluation with verifiers from similar model families therefore risks producing overly optimistic \emph{in silico} success rates for hallucination-based methods, even when the evaluation verifier and the design-time scoring model are nominally different checkpoints.
We view this as \emph{verifier-dependent bias}, which is closely tied to one motivation of our work: a strong score under a single co-folding verifier should not be treated as equivalent to wet-lab truth.
A systematic verifier-score correlation analysis stratified by hallucination versus non-hallucination methods is left as future work; as a complementary retrospective on AI-generated candidate distributions, we additionally evaluate verifier behavior against wet-lab outcomes from RFdiffusion designs in \Cref{appendix:rfdiffusion-retrospective}.

Our generative benchmarks further highlight that practical binder design performance cannot be characterized by a single metric: throughput, per-sequence success rate, and structural diversity are partially decoupled and reveal complementary aspects of model behavior. High per-sample success does not necessarily translate into high yield under realistic resource constraints, while models with lower raw success rates may nevertheless achieve broad coverage of productive regions of the design space. These trade-offs underscore the importance of throughput-aware and diversity-aware evaluation when comparing generative models intended for real-world deployment.

Beyond the analyses presented here, several promising directions emerge for extending the \ourmodel framework. First, exploring combinations of diverse scoring models: integrating scores from a wider range of models, such as energy-based or molecular-dynamics-based metrics, has the potential to enable more precise screening. Second, expanding to broader molecular design applications: the \ourmodel framework could be extended to other areas of molecular design, such as the virtual screening of peptides and small molecules. Third, improving success rates for challenging targets: although \ourmodel introduces more accurate and robust screening metrics, there is still room to improve screening success rates, particularly for notoriously difficult targets.


\newpage
\section*{Impact Statement}
The proposed framework can facilitate more meaningful comparison between emerging generative models and help the community better understand trade-offs between efficiency, diversity, and success under realistic computational constraints.
In the long term, such standardized evaluation protocols may accelerate the development of reliable protein binders for applications in biotechnology.

Potential negative impacts include the misuse of computational protein design and evaluation frameworks for the intentional design or optimization of harmful or toxic viral agents.
Our work focuses exclusively on benchmarking and \emph{in silico} evaluation, and does not provide biological design rules, experimental protocols, or deployment guidance.
We mitigate such risks by emphasizing that all reported results are conditional on specific verification protocols and do not imply biological function or pathogenicity, and we encourage future work to incorporate appropriate ethical oversight and safeguards when applying these methods to biologically hazardous targets.

\bibliography{reference}

@article{chen2025protenix,
  title={Protenix - Advancing Structure Prediction Through a Comprehensive AlphaFold3 Reproduction},
  author={Chen, Xinshi and Zhang, Yuxuan and Lu, Chan and Ma, Wenzhi and Guan, Jiaqi and Gong, Chengyue and Yang, Jincai and Zhang, Hanyu and Zhang, Ke and Wu, Shenghao and Zhou, Kuangqi and Yang, Yanping and Liu, Zhenyu and Wang, Lan and Shi, Bo and Shi, Shaochen and Xiao, Wenzhi},
  year={2025},
  doi = {10.1101/2025.01.08.631967},
  journal = {bioRxiv}
}

@article{gong2025protenix,
  title={Protenix-Mini: Efficient Structure Predictor via Compact Architecture, Few-Step Diffusion and Switchable pLM},
  author={Gong, Chengyue and Chen, Xinshi and Zhang, Yuxuan and Song, Yuxuan and Zhou, Hao and Xiao, Wenzhi},
  journal={arXiv preprint arXiv:2507.11839},
  year={2025}
}

@article{abramson2024accurate,
  title={Accurate structure prediction of biomolecular interactions with AlphaFold 3},
  author={Abramson, Josh and Adler, Jonas and Dunger, Jack and Evans, Richard and Green, Tim and Pritzel, Alexander and Ronneberger, Olaf and Willmore, Lindsay and Ballard, Andrew J and Bambrick, Joshua and others},
  journal={Nature},
  volume={630},
  number={8016},
  pages={493--500},
  year={2024},
  publisher={Nature Publishing Group UK London}
}

@article{wohlwend2024boltz,
  title={Boltz-1 democratizing biomolecular interaction modeling},
  author={Wohlwend, Jeremy and Corso, Gabriele and Passaro, Saro and Reveiz, Mateo and Leidal, Ken and Swiderski, Wojtek and Portnoi, Tally and Chinn, Itamar and Silterra, Jacob and Jaakkola, Tommi and others},
  journal={BioRxiv},
  year={2024}
}

@article{cao2022design,
  title={Design of protein-binding proteins from the target structure alone},
  author={Cao, Longxing and Coventry, Brian and Goreshnik, Inna and Huang, Buwei and Sheffler, William and Park, Joon Sung and Jude, Kevin M and Markovi{\'c}, Iva and Kadam, Rameshwar U and Verschueren, Koen HG and others},
  journal={Nature},
  volume={605},
  number={7910},
  pages={551--560},
  year={2022},
  publisher={Nature Publishing Group UK London}
}

@article{jumper2021highly,
  title={Highly accurate protein structure prediction with AlphaFold},
  author={Jumper, John and Evans, Richard and Pritzel, Alexander and Green, Tim and Figurnov, Michael and Ronneberger, Olaf and Tunyasuvunakool, Kathryn and Bates, Russ and {\v{Z}}{\'\i}dek, Augustin and Potapenko, Anna and others},
  journal={nature},
  volume={596},
  number={7873},
  pages={583--589},
  year={2021},
  publisher={Nature Publishing Group}
}

@article{pacesa2025one,
  title={One-shot design of functional protein binders with BindCraft},
  author={Pacesa, Martin and Nickel, Lennart and Schellhaas, Christian and Schmidt, Joseph and Pyatova, Ekaterina and Kissling, Lucas and Barendse, Patrick and Choudhury, Jagrity and Kapoor, Srajan and Alcaraz-Serna, Ana and others},
  journal={Nature},
  pages={1--10},
  year={2025},
  publisher={Nature Publishing Group UK London}
}

@article{mirdita2022colabfold,
  title={ColabFold: making protein folding accessible to all},
  author={Mirdita, Milot and Sch{\"u}tze, Konstantin and Moriwaki, Yoshitaka and Heo, Lim and Ovchinnikov, Sergey and Steinegger, Martin},
  journal={Nature methods},
  volume={19},
  number={6},
  pages={679--682},
  year={2022},
  publisher={Nature Publishing Group US New York}
}

@article{evans2021protein,
  title={Protein complex prediction with AlphaFold-Multimer},
  author={Evans, Richard and O’Neill, Michael and Pritzel, Alexander and Antropova, Natasha and Senior, Andrew and Green, Tim and {\v{Z}}{\'\i}dek, Augustin and Bates, Russ and Blackwell, Sam and Yim, Jason and others},
  journal={biorxiv},
  pages={2021--10},
  year={2021},
  publisher={Cold Spring Harbor Laboratory}
}

@article{cho2025boltzdesign1,
  title={Boltzdesign1: Inverting all-atom structure prediction model for generalized biomolecular binder design},
  author={Cho, Yehlin and Pacesa, Martin and Zhang, Zhidian and Correia, Bruno E and Ovchinnikov, Sergey},
  journal={bioRxiv},
  pages={2025--04},
  year={2025},
  publisher={Cold Spring Harbor Laboratory}
}

@article{chai2024chai,
  title={Chai-1: Decoding the molecular interactions of life},
  author={ChaiDiscovery and Boitreaud, Jacques and Dent, Jack and McPartlon, Matthew and Meier, Joshua and Reis, Vinicius and Rogozhonikov, Alex and Wu, Kevin},
  journal={BioRxiv},
  pages={2024--10},
  year={2024},
  publisher={Cold Spring Harbor Laboratory}
}

@article{zambaldi2024novo,
  title={De novo design of high-affinity protein binders with AlphaProteo},
  author={Zambaldi, Vinicius and La, David and Chu, Alexander E and Patani, Harshnira and Danson, Amy E and Kwan, Tristan OC and Frerix, Thomas and Schneider, Rosalia G and Saxton, David and Thillaisundaram, Ashok and others},
  journal={arXiv preprint arXiv:2409.08022},
  year={2024}
}

@article{watson2023novo,
  title={De novo design of protein structure and function with RFdiffusion},
  author={Watson, Joseph L and Juergens, David and Bennett, Nathaniel R and Trippe, Brian L and Yim, Jason and Eisenach, Helen E and Ahern, Woody and Borst, Andrew J and Ragotte, Robert J and Milles, Lukas F and others},
  journal={Nature},
  volume={620},
  number={7976},
  pages={1089--1100},
  year={2023},
  publisher={Nature Publishing Group UK London}
}

@article{lin2023evolutionary,
  title={Evolutionary-scale prediction of atomic-level protein structure with a language model},
  author={Lin, Zeming and Akin, Halil and Rao, Roshan and Hie, Brian and Zhu, Zhongkai and Lu, Wenting and Smetanin, Nikita and Verkuil, Robert and Kabeli, Ori and Shmueli, Yaniv and others},
  journal={Science},
  volume={379},
  number={6637},
  pages={1123--1130},
  year={2023},
  publisher={American Association for the Advancement of Science}
}

@article{asimit2017robust,
  title={Robust and Pareto optimality of insurance contracts},
  author={Asimit, Alexandru V and Bignozzi, Valeria and Cheung, Ka Chun and Hu, Junlei and Kim, Eun-Seok},
  journal={European Journal of Operational Research},
  volume={262},
  number={2},
  pages={720--732},
  year={2017},
  publisher={Elsevier}
}

@article{krishna2024generalized,
  title={Generalized biomolecular modeling and design with RoseTTAFold All-Atom},
  author={Krishna, Rohith and Wang, Jue and Ahern, Woody and Sturmfels, Pascal and Venkatesh, Preetham and Kalvet, Indrek and Lee, Gyu Rie and Morey-Burrows, Felix S and Anishchenko, Ivan and Humphreys, Ian R and others},
  journal={Science},
  volume={384},
  number={6693},
  pages={eadl2528},
  year={2024},
  publisher={American Association for the Advancement of Science}
}

@article{van2024fast,
  title={Fast and accurate protein structure search with Foldseek},
  author={Van Kempen, Michel and Kim, Stephanie S and Tumescheit, Charlotte and Mirdita, Milot and Lee, Jeongjae and Gilchrist, Cameron LM and S{\"o}ding, Johannes and Steinegger, Martin},
  journal={Nature biotechnology},
  volume={42},
  number={2},
  pages={243--246},
  year={2024},
  publisher={Nature Publishing Group US New York}
}

@article{biogeometry2025geoflow,
  title={GeoFlow-V2: A Unified Atomic Diffusion Model for Protein Structure Prediction and De Novo Design},
  author={BioGeometry},
  journal={bioRxiv},
  pages={2025--05},
  year={2025},
  publisher={Cold Spring Harbor Laboratory}
}

@article{bridgland2025latent,
  title={Latent-X: An Atom-level Frontier Model for De Novo Protein Binder Design},
  author={Bridgland, Alex and Crabb{\'e}, Jonathan and Kenlay, Henry and Pretorius, Daniella and Schmon, Sebastian M and Hilmkil, Agrin and Bartke-Croughan, Rebecca and Rombach, Robin and Flashman, Michael and Matteson, Tomas and others},
  journal={arXiv e-prints},
  pages={arXiv--2507},
  year={2025}
}

@article{chai2025zero,
  title={Zero-shot antibody design in a 24-well plate},
  author={ChaiDiscovery and Boitreaud, Jacques and Dent, Jack and Geisz, Danny and McPartlon, Matthew and Meier, Joshua and Qiao, Zhuoran and Rogozhnikov, Alex and Rollins, Nathan and Wollenhaupt, Paul and others},
  journal={bioRxiv},
  pages={2025--07},
  year={2025},
  publisher={Cold Spring Harbor Laboratory}
}

@article{butcher2025novo,
  title={De novo Design of All-atom Biomolecular Interactions with RFdiffusion3},
  author={Butcher, Jasper Kenneth Veje and Krishna, Rohith and Mitra, Raktim and Brent, Rafael Isaac and Li, Yanjing and Corley, Nathaniel and Kim, Paul and Funk, Jonathan and Mathis, Simon Valentin and Salike, Saman and others},
  journal={bioRxiv},
  pages={2025--09},
  year={2025},
  publisher={Cold Spring Harbor Laboratory}
}

@article{kortemme2002simple,
  title={A simple physical model for binding energy hot spots in protein--protein complexes},
  author={Kortemme, Tanja and Baker, David},
  journal={Proceedings of the National Academy of Sciences},
  volume={99},
  number={22},
  pages={14116--14121},
  year={2002},
  publisher={National Academy of Sciences}
}

@article{baek2021accurate,
  title={Accurate prediction of protein structures and interactions using a three-track neural network},
  author={Baek, Minkyung and DiMaio, Frank and Anishchenko, Ivan and Dauparas, Justas and Ovchinnikov, Sergey and Lee, Gyu Rie and Wang, Jue and Cong, Qian and Kinch, Lisa N and Schaeffer, R Dustin and others},
  journal={Science},
  volume={373},
  number={6557},
  pages={871--876},
  year={2021},
  publisher={American Association for the Advancement of Science}
}

@article{yang2020improved,
  title={Improved protein structure prediction using predicted interresidue orientations},
  author={Yang, Jianyi and Anishchenko, Ivan and Park, Hahnbeom and Peng, Zhenling and Ovchinnikov, Sergey and Baker, David},
  journal={Proceedings of the National Academy of Sciences},
  volume={117},
  number={3},
  pages={1496--1503},
  year={2020},
  publisher={National Academy of Sciences}
}

@article {ren2025pxdesign,
	author = {Protenix Team and Ren, Milong and Sun, Jinyuan and Guan, Jiaqi and Liu, Cong and Gong, Chengyue and Wang, Yuzhe and Wang, Lan and Cai, Qixu and Chen, Xinshi and Xiao, Wenzhi},
	title = {PXDesign: Fast, Modular, and Accurate De Novo Design of Protein Binders},
	elocation-id = {2025.08.15.670450},
	year = {2025},
	doi = {10.1101/2025.08.15.670450},
	publisher = {Cold Spring Harbor Laboratory},
	URL = {https://www.biorxiv.org/content/early/2025/08/16/2025.08.15.670450},
	eprint = {https://www.biorxiv.org/content/early/2025/08/16/2025.08.15.670450.full.pdf},
	journal = {bioRxiv}
}

@misc{zhang2025odesignworldmodelbiomolecular,
      title={ODesign: A World Model for Biomolecular Interaction Design}, 
      author={Odin Zhang and Xujun Zhang and Haitao Lin and Cheng Tan and Qinghan Wang and Yuanle Mo and Qiantai Feng and Gang Du and Yuntao Yu and Zichang Jin and Ziyi You and Peicong Lin and Yijie Zhang and Yuyang Tao and Shicheng Chen and Jack Xiaoyu Chen and Chenqing Hua and Weibo Zhao and Runze Ma and Yunpeng Xia and Kejun Ying and Jun Li and Yundian Zeng and Lijun Lang and Peichen Pan and Hanqun Cao and Zihao Song and Bo Qiang and Jiaqi Wang and Pengfei Ji and Lei Bai and Jian Zhang and Chang-yu Hsieh and Pheng Ann Heng and Siqi Sun and Tingjun Hou and Shuangjia Zheng},
      year={2025},
      eprint={2510.22304},
      archivePrefix={arXiv},
      primaryClass={q-bio.BM},
      url={https://arxiv.org/abs/2510.22304}, 
}

@article {Stark2025BoltzGen,
	author = {Stark, Hannes and Faltings, Felix and Choi, MinGyu and Xie, Yuxin and Hur, Eunsu and O{\textquoteright}Donnell, Timothy and Bushuiev, Anton and U{\c c}ar, Talip and Passaro, Saro and Mao, Weian and Reveiz, Mateo and Bushuiev, Roman and Pluskal, Tom{\'a}{\v s} and Sivic, Josef and Kreis, Karsten and Vahdat, Arash and Ray, Shamayeeta and Goldstein, Jonathan T. and Savinov, Andrew and Hambalek, Jacob A. and Gupta, Anshika and Taquiri-Diaz, Diego A. and Zhang, Yaotian and Hatstat, A. Katherine and Arada, Angelika and Kim, Nam Hyeong and Tackie-Yarboi, Ethel and Boselli, Dylan and Schnaider, Lee and Liu, Chang C. and Li, Gene-Wei and Hnisz, Denes and Sabatini, David M. and DeGrado, William F. and Wohlwend, Jeremy and Corso, Gabriele and Barzilay, Regina and Jaakkola, Tommi},
	title = {BoltzGen: Toward Universal Binder Design},
	elocation-id = {2025.11.20.689494},
	year = {2025},
	doi = {10.1101/2025.11.20.689494},
	publisher = {Cold Spring Harbor Laboratory},
	URL = {https://www.biorxiv.org/content/early/2025/11/24/2025.11.20.689494},
	eprint = {https://www.biorxiv.org/content/early/2025/11/24/2025.11.20.689494.full.pdf},
	journal = {bioRxiv}
}

@article{passaro2025boltz2,
  author = {Passaro, Saro and Corso, Gabriele and Wohlwend, Jeremy and Reveiz, Mateo and Thaler, Stephan and Somnath, Vignesh Ram and Getz, Noah and Portnoi, Tally and Roy, Julien and Stark, Hannes and Kwabi-Addo, David and Beaini, Dominique and Jaakkola, Tommi and Barzilay, Regina},
  title = {Boltz-2: Towards Accurate and Efficient Binding Affinity Prediction},
  year = {2025},
  doi = {10.1101/2025.06.14.659707},
  journal = {bioRxiv}
}

@misc{qu2025seedproteoaccuratenovoallatom,
      title={SeedProteo: Accurate De Novo All-Atom Design of Protein Binders}, 
      author={Wei Qu and Yiming Ma and Fei Ye and Chan Lu and Yi Zhou and Kexin Zhang and Lan Wang and Minrui Gui and Quanquan Gu},
      year={2025},
      eprint={2512.24192},
      archivePrefix={arXiv},
      primaryClass={q-bio.BM},
      url={https://arxiv.org/abs/2512.24192}, 
}

@article {chu2023allatom,
    author = {Alexander E. Chu and Lucy Cheng and Gina El Nesr and Minkai Xu and Po-Ssu Huang},
    title = {An all-atom protein generative model},
    year = {2023},
    doi = {10.1101/2023.05.24.542194},
    URL = {https://www.biorxiv.org/content/early/2023/05/25/2023.05.24.542194},
    journal = {bioRxiv}
}
\bibliographystyle{icml2026}

\newpage
\appendix
\onecolumn
\section{Benchmark Details}
\label{appendix:benchmark}

\begin{table}
\centering
\small
\begin{tabularx}{\textwidth}{l l X X l l}
\toprule
\textbf{Target} & \textbf{PDB ID} &  \textbf{Crop} & \textbf{Hotspot} &
\textbf{Natural binder} & 
\textbf{Binder length} \\
\midrule
BHRF1   & \texttt{2wh6}  & A2--158                             & A65, A74, A77, \newline A82, A85, A93             & BH3 helix             & 80--120 \\
SC2RBD  & \texttt{6m0j}  & E333--526                           & E485, E489, \newline E494, E500, \newline E505             & ACE2 receptor         & 80--120 \\
IL-7RA  & \texttt{3di3}  & B17--209                            & B58, B80, B139                            & IL-7                  & 50--120 \\
PD-L1   & \texttt{5o45}  & A17--132                            & A56, A115, \newline A123                           & PD-1                  & 50--120 \\
TrkA    & \texttt{1www}  & X282--382                           & X294, X296, \newline X333                          & Nerve growth factor   & 50--120 \\
Insulin & \texttt{4zxb}  & E6--155                             & E64, E88, E96                             & Insulin receptor      & 40--120 \\
H1      & \texttt{5vli}  & A1--45,\newline A47--50, \newline A76--80, \newline A107--111, \newline A258--322, \newline B1--68, \newline B80--170 & B21, B45, B52                             & None                  & 40--120 \\
VEGF-A  & \texttt{1bj1}  & V14--107, \newline W14--107                  & W81, W83, \newline W91                             & VEGFR1, VEGFR2        & 50--140 \\
IL-17A  & \texttt{4hsa}  & A17--29, \newline A41--131, \newline B19--127                  & A94, A116, B67                            & IL-17 receptor alpha  & 50--140 \\
TNF$\alpha$ & \texttt{1tnf}  & A12--157, \newline B12--157, \newline C12--157      & A113, C73                                 & None                  & 50--120 \\
\bottomrule
\end{tabularx}
\caption{Overview of benchmark targets for protein binder design. Residue ranges and hotspot annotations are based on known structures and wet-lab characterization.}
\label{tab:design-targets}
\end{table}

\subsection{Task Definition}
\label{target_infomation}
Each binder design task in our benchmark is specified by the following components:

\begin{itemize}[left=5mm]
    \item {\bf Target}: The biological macromolecule that the designed binder is intended to bind, such as a receptor, enzyme, or viral protein.
    
    \item {\bf Crop}: Annotations specifying the target chains and residue ranges used as input to the design model. These typically correspond to solvent-exposed regions of the target surface that are accessible for binding. Since the computational cost of generative models and structure predictors often scales with the number of residues $L$, appropriate cropping improves both runtime efficiency and focus.
    
    \item {\bf Hotspot}: A small subset of residues on the target surface believed to play a critical role in molecular recognition or binding energetics. These are typically derived from known protein–protein complexes, mutagenesis studies, or epitope mapping, and serve as anchors for binder placement.
    
    \item {\bf Binder Length Range}: Specifies the allowed length interval for generated binder sequences. This constraint reflects practical considerations: shorter binders are easier to express, purify, and characterize, while longer sequences may form more stable folds or allow for larger interaction surfaces.
\end{itemize}

The selection of targets, cropping strategies, hotspot residues, and binder length ranges follows established conventions from prior work such as AlphaProteo~\citep{zambaldi2024novo}. A complete overview of the benchmark targets is provided in Table~\ref{tab:design-targets}. We made two minor modifications to the crop definitions compared to AlphaProteo: (1) For target H1, residue 46 includes an insertion code, which is not supported by most baseline models. This residue was excluded from the crop. (2) For target IL-17A, residues A30-A40 are unresolved in the crystal structure and were omitted.

\subsection{Benchmark Preprocessing and Caveats}

\paragraph{Preprocessing of inference outputs.}
To ensure consistency and reproducibility across all methods, 
we harmonized the outputs of different design methods to enable unified downstream structure prediction and filtering:
\begin{itemize}[left=5mm]
    \item For generative models' outputs, we explicitly marker binder chain ids to ensure compatibility with Protenix filters, which utilize target MSAs. 
    \item For all methods, we renumbered residue IDs in the target chains to match those in the original uncropped PDB structures. This ensures correct relative positional encoding for structure predictors.
\end{itemize}

\paragraph{Tool-specific limitations and failure cases.}
We observed several tool-specific limitations that impact compatibility or performance in certain scenarios. For example, BoltzDesign does not properly support cropped residue ranges: even when provided with discontinuous residue indices, it outputs continuous structures that ignore crop boundaries. Additionally, for challenging targets such as TNF$\alpha$, we found that Protenix occasionally produces binder structures that appear disordered or structurally implausible. This typically occurs when the input sequence—e.g., one generated by ProteinMPNN—contains excessive alanines, leading the structure predictor to interpret the region as intrinsically disordered or low-confidence.

\subsection{More benchmark results}


\begin{table}[t]
\caption{Secondary Structure Ratio and Radius of Gyration}
\label{tab:ss-rg}
\begin{center}
\begin{tabular}{m{3.8cm}<{\centering}|
                m{2.0cm}<{\centering}
                m{3.0cm}<{\centering}}
\toprule[1.5pt]
\textbf{Method} 
& \textbf{Rg} 
& $\boldsymbol{\alpha/\beta}$ \\
& & \\
\hline
Protpardelle-1c & 13.49 & 24.2\%:31.4\% \\
RFDiffusion-3   & 13.09 & 70.4\%:8.2\%  \\
Odesign         & 12.97 & 46.5\%:19.3\% \\
Boltzgen        & 12.43 & 36.1\%:24.6\% \\
PXDesign        & 13.00 & 81.8\%:3.5\%  \\
BindCraft       & 13.37 & 72.4\%:2.3\%  \\
BoltzDesign-1   & 14.81 & 42.0\%:19.1\% \\
\bottomrule[1.5pt]
\end{tabular}
\end{center}
\end{table}



\textbf{Confidence-free property.}  \hspace{0.2cm} We also compared the properties of binders designed by different methods, including the proportion of various secondary structures (e.g., $\alpha$/$\beta$) and the radius of gyration (Rg) of the designed binders, as summarized in \Cref{tab:ss-rg}. Overall, different methods exhibit distinct preferences in secondary structure composition and compactness, reflecting their inherent design biases.

\begin{figure}[htbp]  
  \centering  
  \begin{subfigure}[b]{0.49\textwidth}  
    \centering 
    \includegraphics[width=\textwidth]{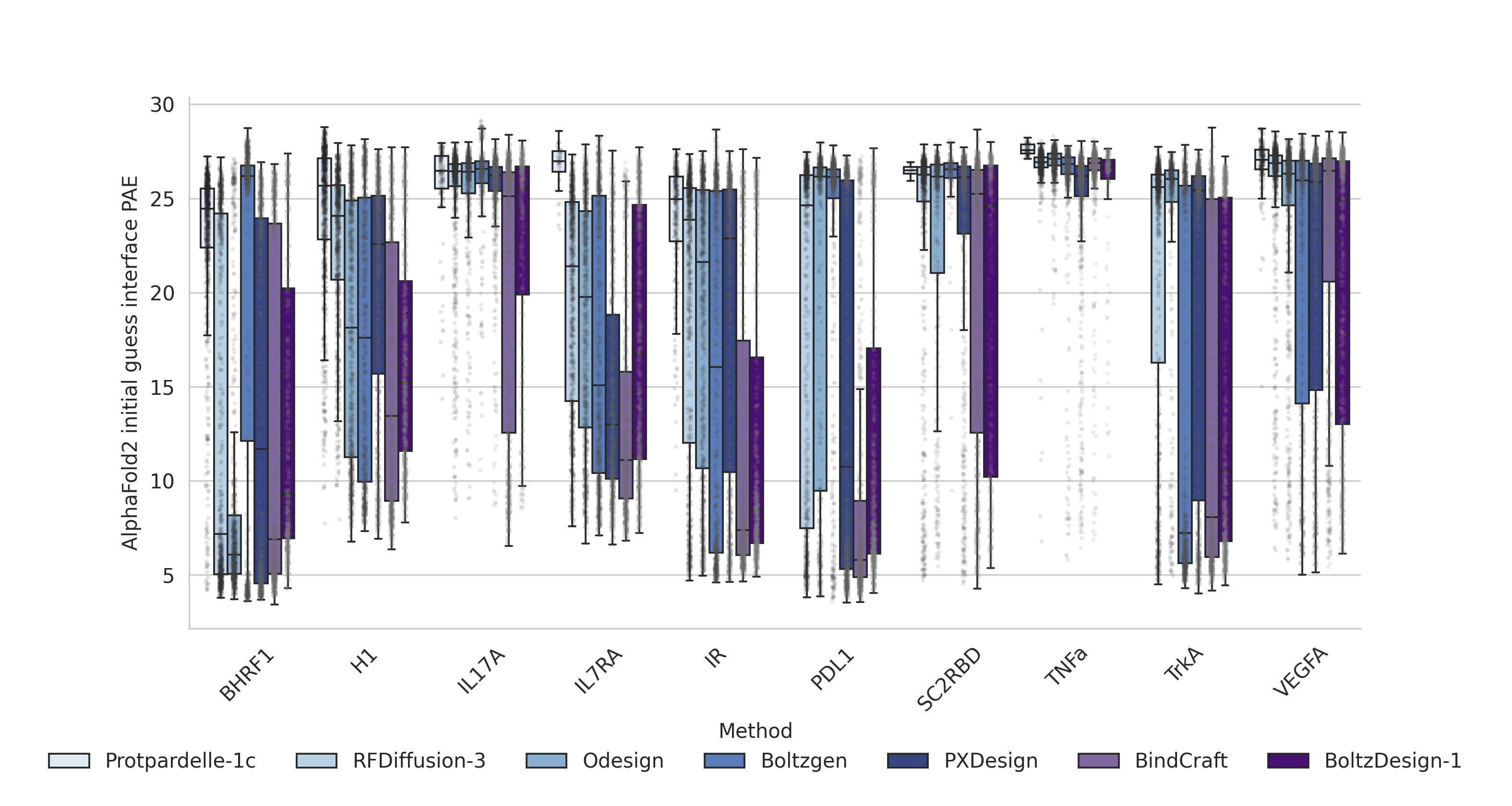}
    \caption{All-alpha binder performance on AF2 metric.}  
    \label{fig:a}  
  \end{subfigure}
  \begin{subfigure}[b]{0.49\textwidth}  
    \centering
    \includegraphics[width=\textwidth]{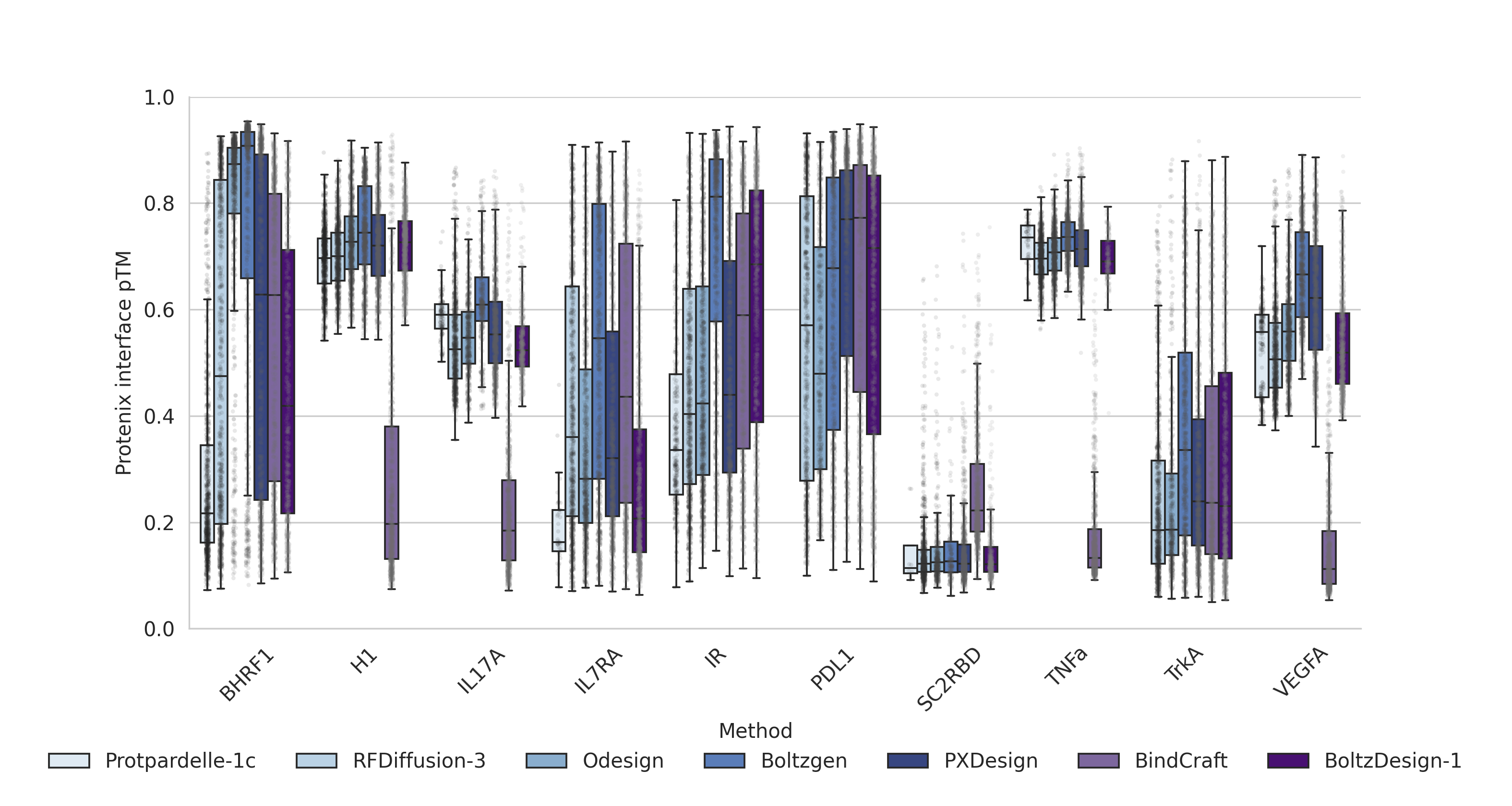} 
    \caption{All-alpha binder performance on Protenix metric.} 
    \label{fig:b}
  \end{subfigure}

  \begin{subfigure}[b]{0.49\textwidth} 
    \centering
    \includegraphics[width=\textwidth]{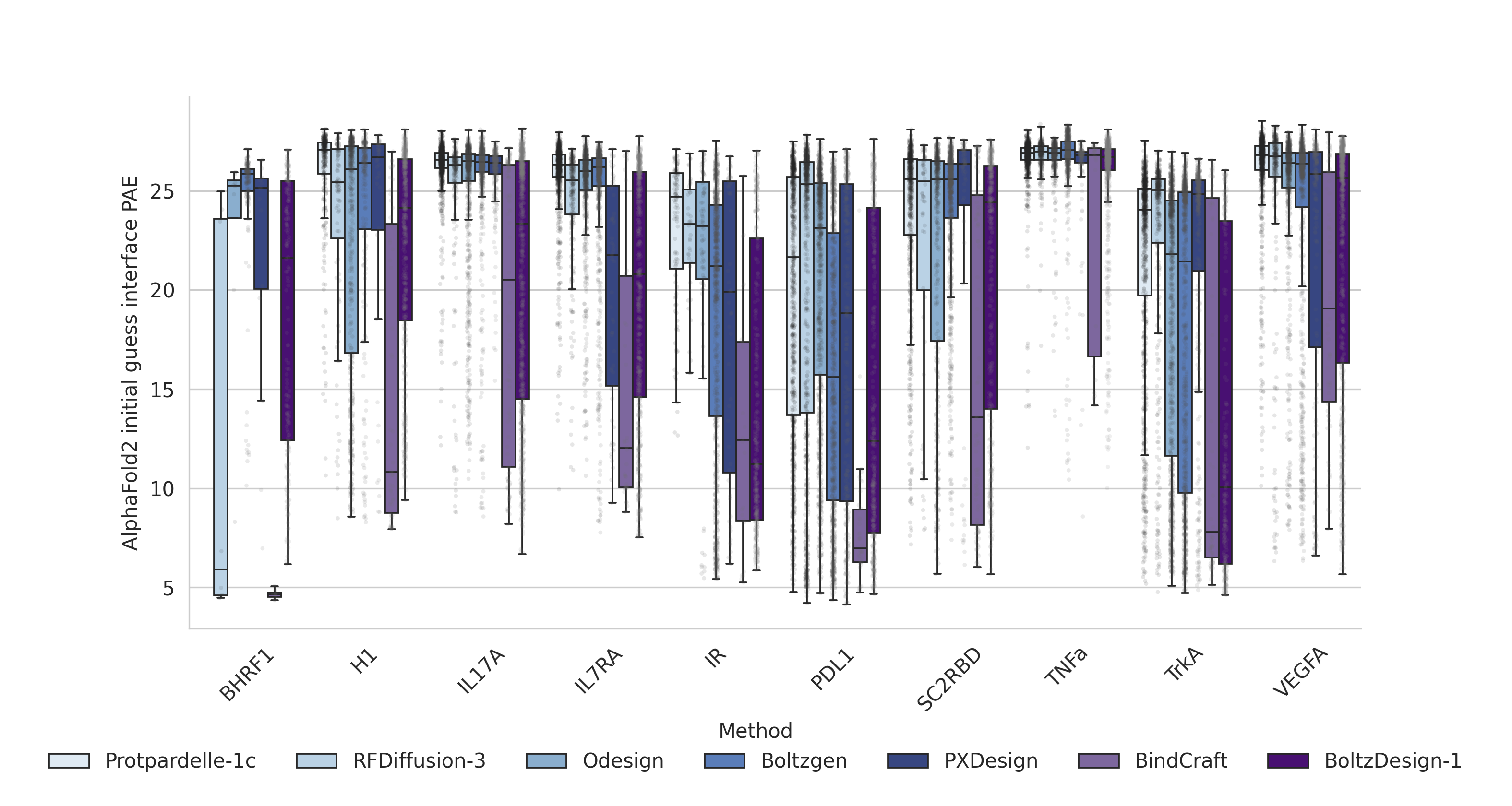}
    \caption{Mainly-beta binder performance on AF2 metric.} 
    \label{fig:c}
  \end{subfigure}
  \begin{subfigure}[b]{0.50\textwidth}  
    \centering
    \includegraphics[width=\textwidth]{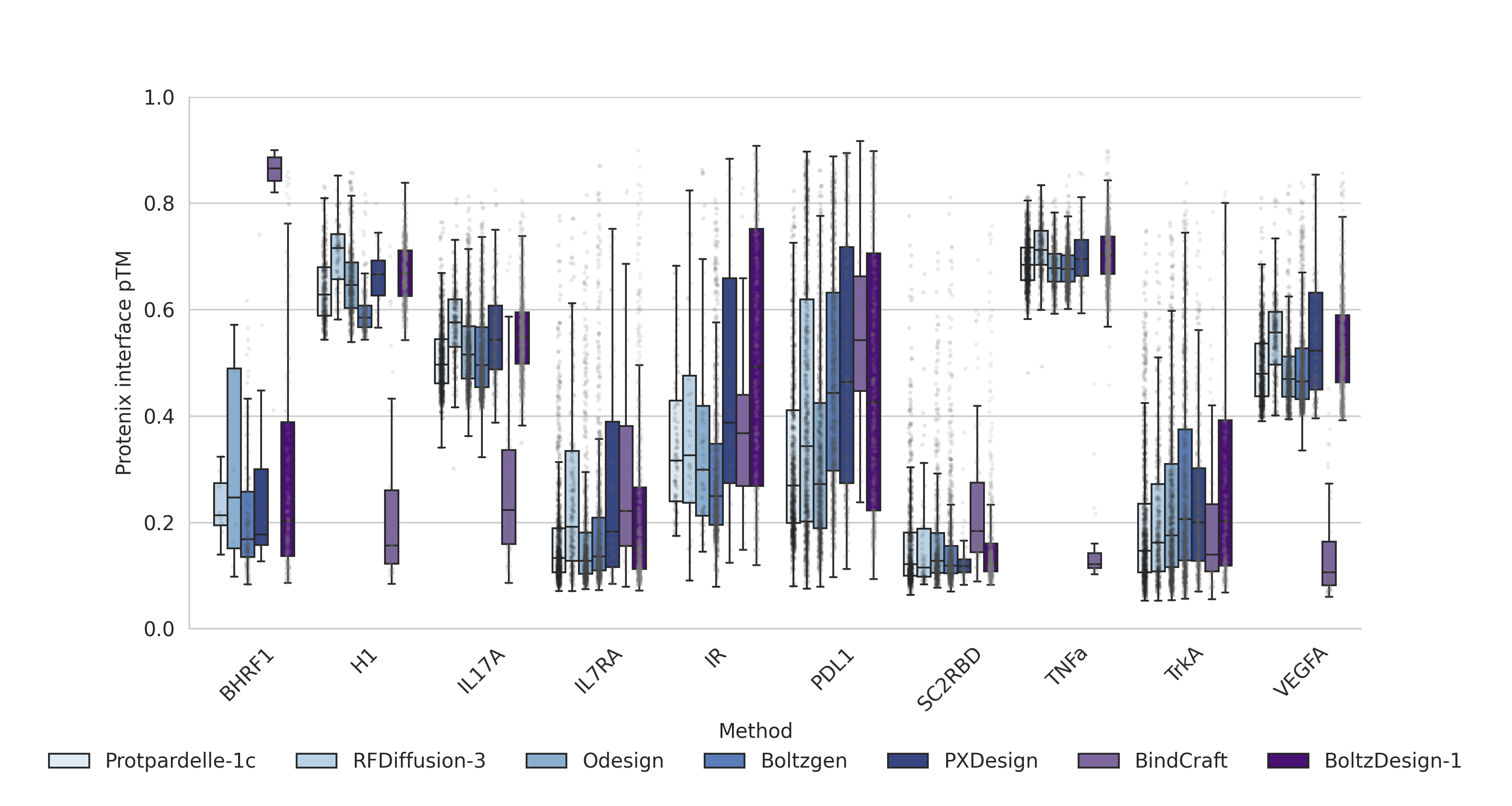}  
    \caption{Mainly-beta binder performance on Protenix metric.} 
    \label{fig:d}
  \end{subfigure}

  \caption{Performance of binders with different secondary structures designed by various methods on AlphaFold2- and Protenix-based metrics.}
  \label{fig:alpha_beta_binder}  
\end{figure}

\textbf{Confidence by secondary structures.}
\label{appendix:other_results}
We show the performance of binders designed by different methods under different secondary structure proportions in \Cref{fig:alpha_beta_binder}. First, we used DSSP to annotate the designed binders for secondary structure, and we divided the designed binders into two categories: all-alpha (beta fraction $<$ $1\%$) and mainly-beta (beta fraction $>$ $30\%$). We found that for all currently designed methods, across almost all targets, whether on AF2- or Protenix-based metrics, the performance of all-alpha-designed binders is better than that of mainly-beta-designed binders. Improving the generation of high-quality, beta-sheet-containing binders is a potential direction for future computational model development.

\subsection{Generative benchmark results with Protenix-Mini as verifier}
\label{app:protenix-mini-benchmark}

In the main paper, we report generative benchmarking results using AF2-IG-Easy as the primary verifier for throughput-aligned comparisons across methods. Here we provide an additional robustness check by re-evaluating the same set of generated binders with \textbf{Protenix-Mini} as an alternative verifier, and report both the \textbf{success rate} and the \textbf{cluster pass rate}.

Protenix-Mini applies a conservative set  when converting its predicted structure/confidence signals into a binary pass/fail decision. As a result, absolute success rates are substantially lower than those obtained under AF2-IG-Easy, and several difficult targets exhibit near-zero pass rates across most methods.

\begin{figure}[t]
    \centering
    \includegraphics[width=0.6\linewidth]{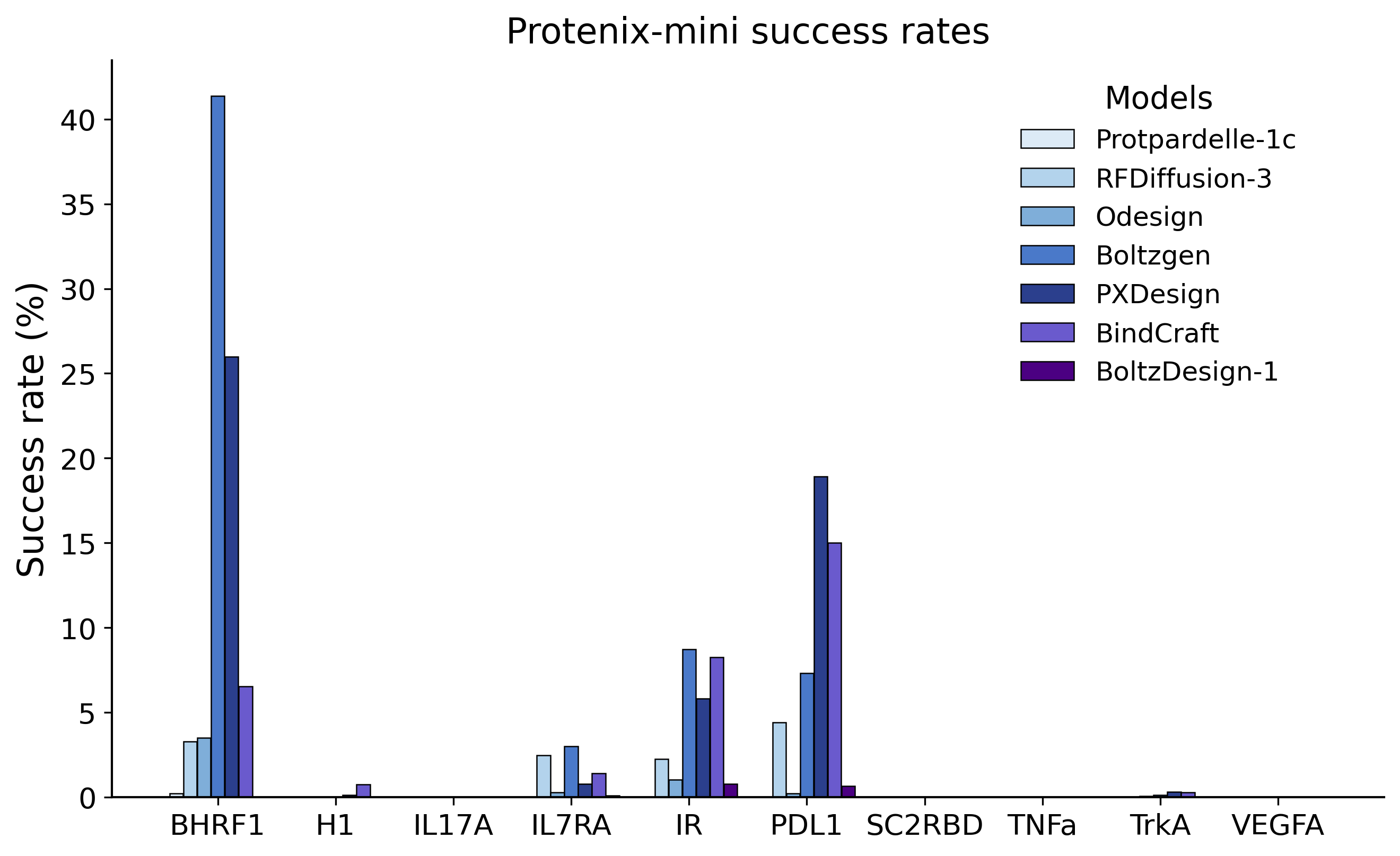}
    \caption{\textbf{Protenix-Mini sequence-level success rates.} Fraction of generated binder sequences passing the Protenix-Mini filter for each target. Absolute success rates are lower than AF2-IG-Easy due to the conservative confidence scores of Protenix-Mini.}
    \label{fig:ptxmini-sr}
\end{figure}
\begin{figure}[t]
    \centering
    \includegraphics[width=\linewidth]{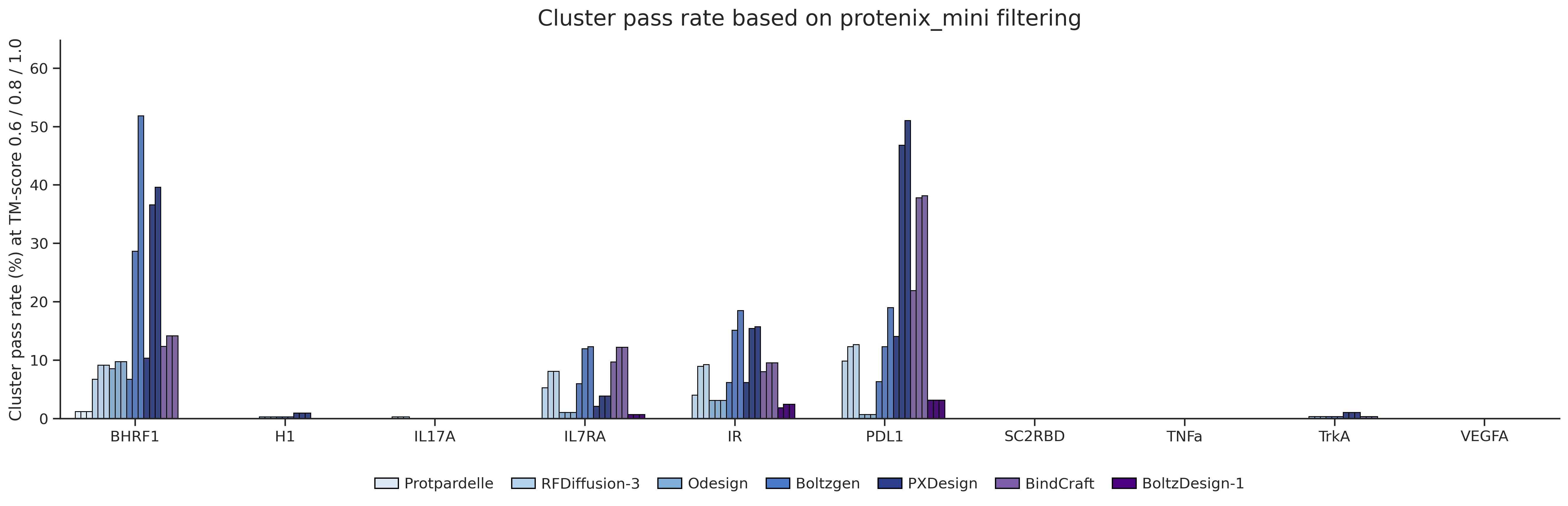}
    \caption{\textbf{Protenix-Mini cluster pass rates.} Diversity-adjusted success rate computed as the number of unique structural clusters among \emph{passed} designs (clustered at TM-score thresholds 0.6/0.8/1.0) divided by the total number of generated backbones. This metric captures both filter success and diversity among successful designs.}
    \label{fig:ptxmini-cpr}
\end{figure}

Figure~\ref{fig:ptxmini-sr} reports the \textbf{Protenix-Mini success rate}, defined as the fraction of generated binder sequences that pass the Protenix-Mini filter for each target. Figure~\ref{fig:ptxmini-cpr} reports the corresponding \textbf{cluster pass rate} (diversity-adjusted success rate), computed as the number of unique structural clusters among \emph{passed} designs (clustered by TM-score thresholds 0.6/0.8/1.0) normalized by the total number of generated backbones. The cluster pass rate therefore captures both filter success and structural diversity among successful designs.

Across targets where Protenix-Mini yields non-trivial acceptance (e.g., BHRF1, PDL1, IR, IL7RA), methods that achieve higher success rates generally also achieve higher cluster pass rates, indicating that the relative trends observed under AF2-IG-Easy are partially preserved when switching to a stricter verifier. At the same time, the sharp reduction in absolute acceptance highlights that verifier choice and default cutoffs can materially affect reported success rates.

\subsection{Diversity across TM-score thresholds}
\label{app:diversity-other-tm}

\begin{figure}[t]
    \centering
    \includegraphics[width=\linewidth]{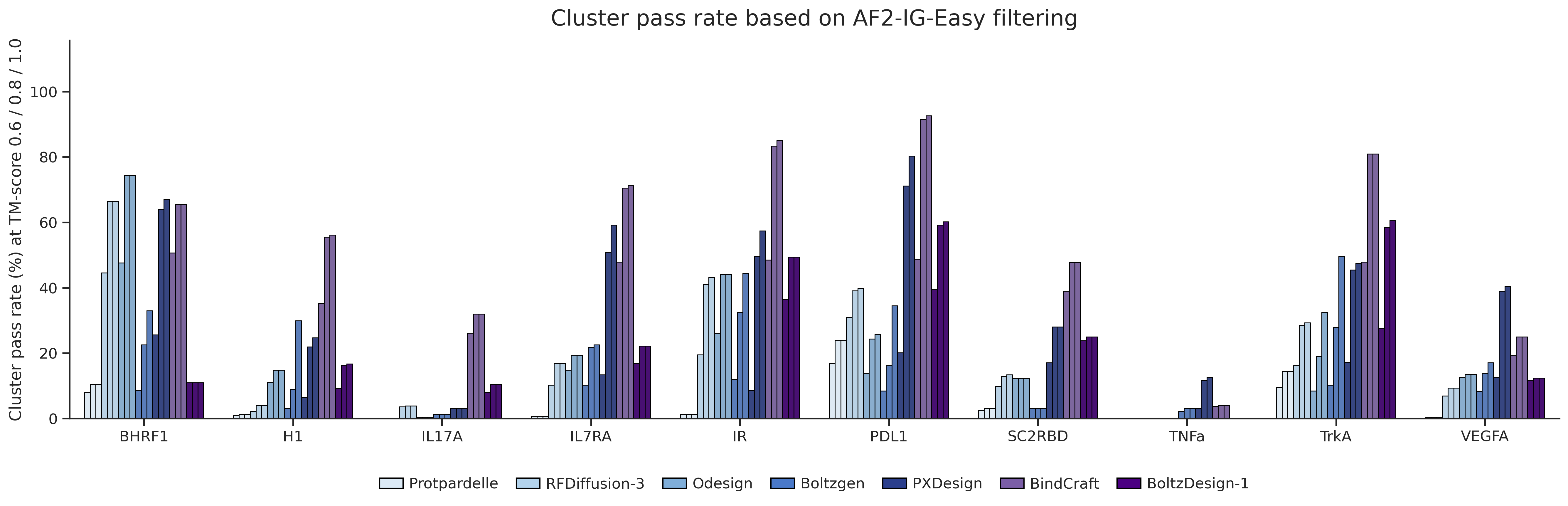}
    \caption{\textbf{Binder structural diversity benchmark across TM-score thresholds.} Diversity-adjusted cluster pass rate after AF2-IG-Easy filtering at TM-score thresholds $0.6$, $0.8$, and $1.0$ (left to right). The leftmost panel matches \Cref{fig:binder_diversity} in the main text.}
    \label{fig:binder_diversity_full}
\end{figure}

\Cref{fig:binder_diversity_full} reports the diversity-adjusted cluster pass rate at multiple TM-score similarity thresholds ($0.6$, $0.8$, and $1.0$), providing a more complete view of structural diversity than the single-threshold main-text figure.

\subsection{Metric details}
Here, we present the distributions of performance metrics for binders designed by various methods against different targets. These metrics include ipAE and plddt from AF2-IG, and iptm and ptm from Protenix.

\begin{figure}[htbp]
    \centering
    \includegraphics[width=0.90\textwidth]{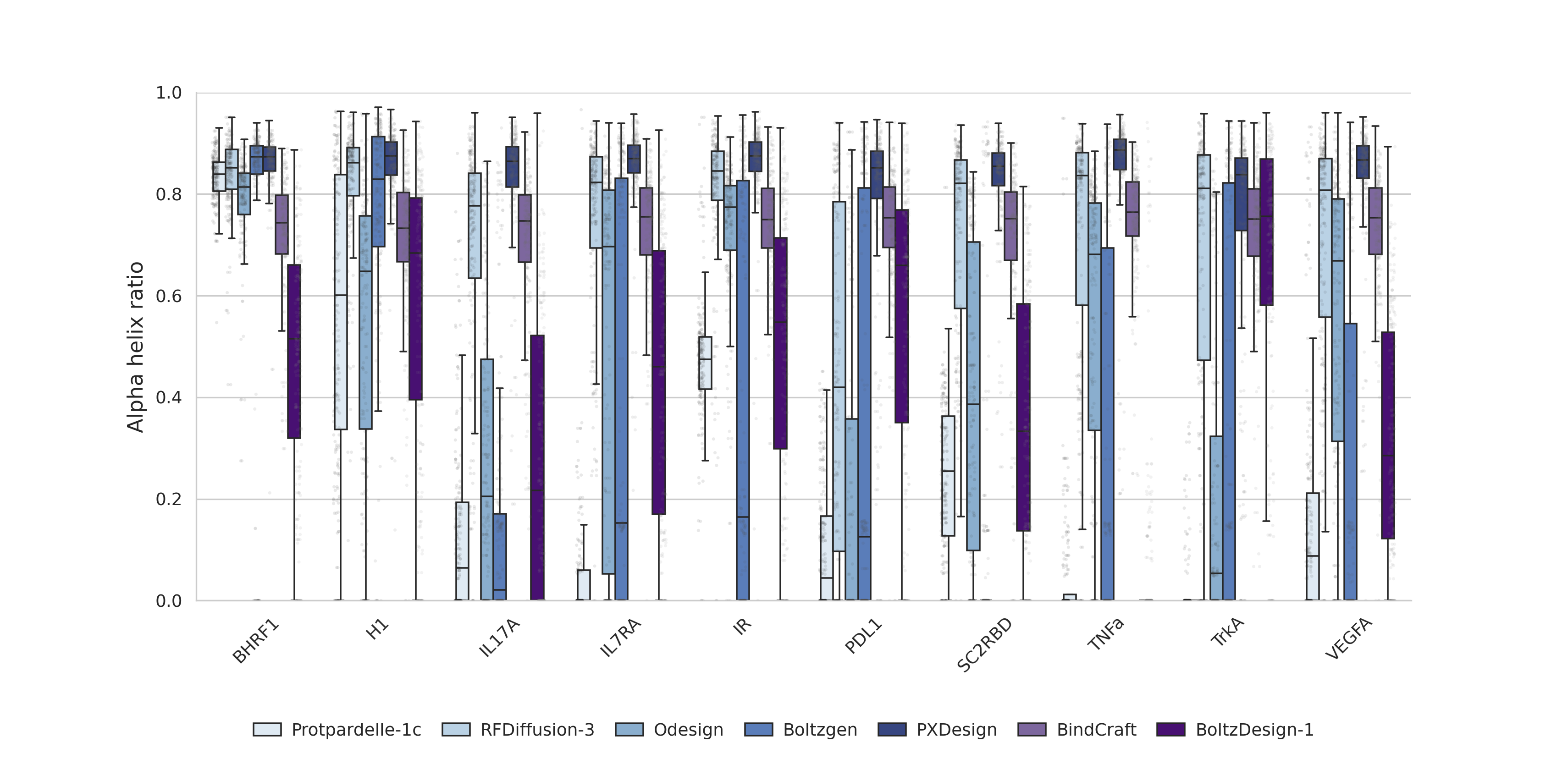}
    \caption{Alpha-helix ratio on different targets across various methods.}
    \label{fig:benchmark_alpha}
\end{figure}

\begin{figure}[htbp]
    \centering
    \includegraphics[width=0.90\textwidth]{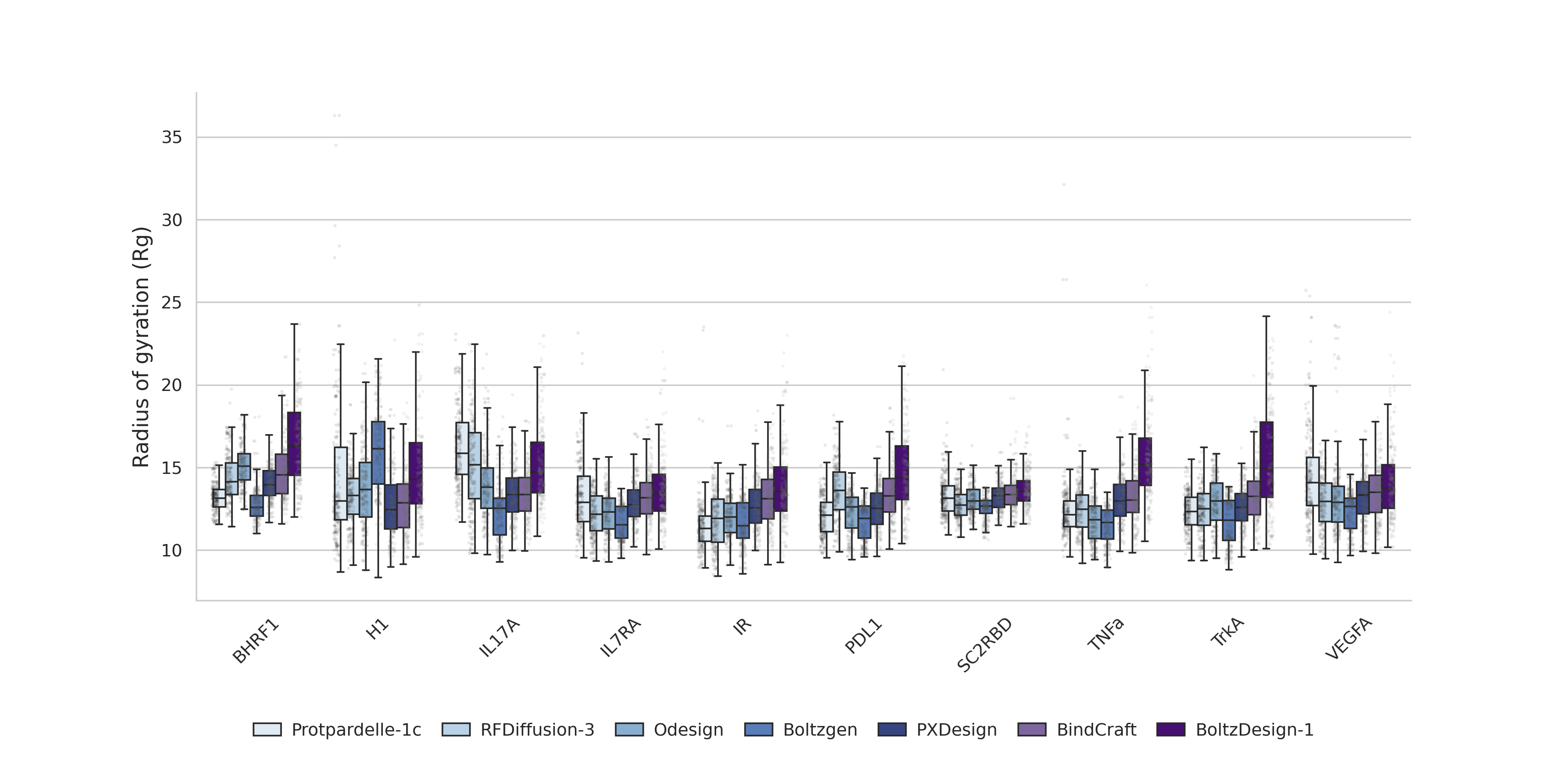}
    \caption{Reference ratio of gyration radius on different targets across various methods.}
    \label{fig:benchmark_ref_ratio}
\end{figure}

\begin{figure}[htbp]
    \centering
    \includegraphics[width=0.90\textwidth]{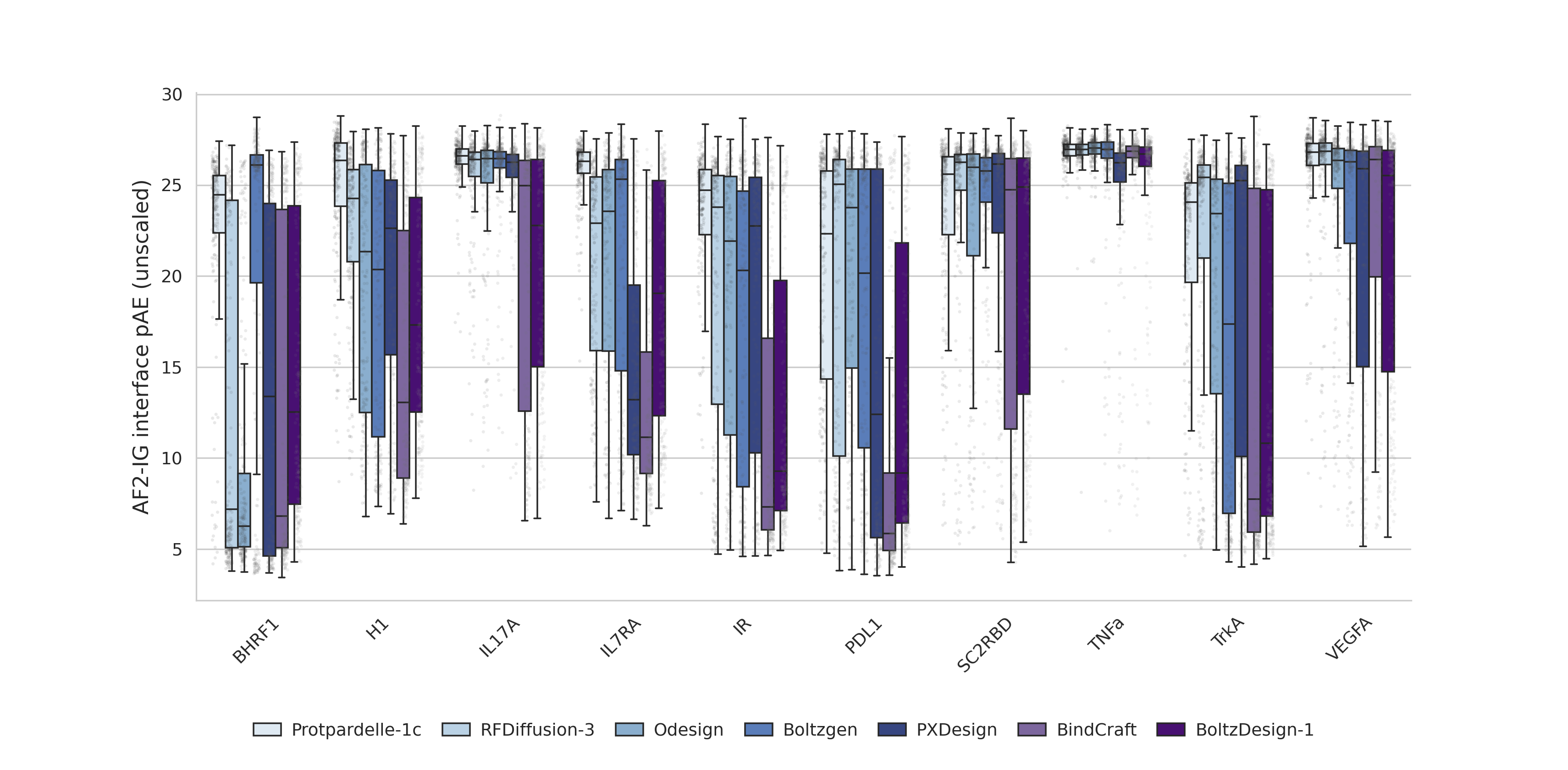}
    \caption{AlphaFold2 initial guess interface pAE on different targets across various methods.}
    \label{fig:benchmark_af2_ipae}
\end{figure}

\begin{figure}[htbp]
    \centering
    \includegraphics[width=0.90\textwidth]{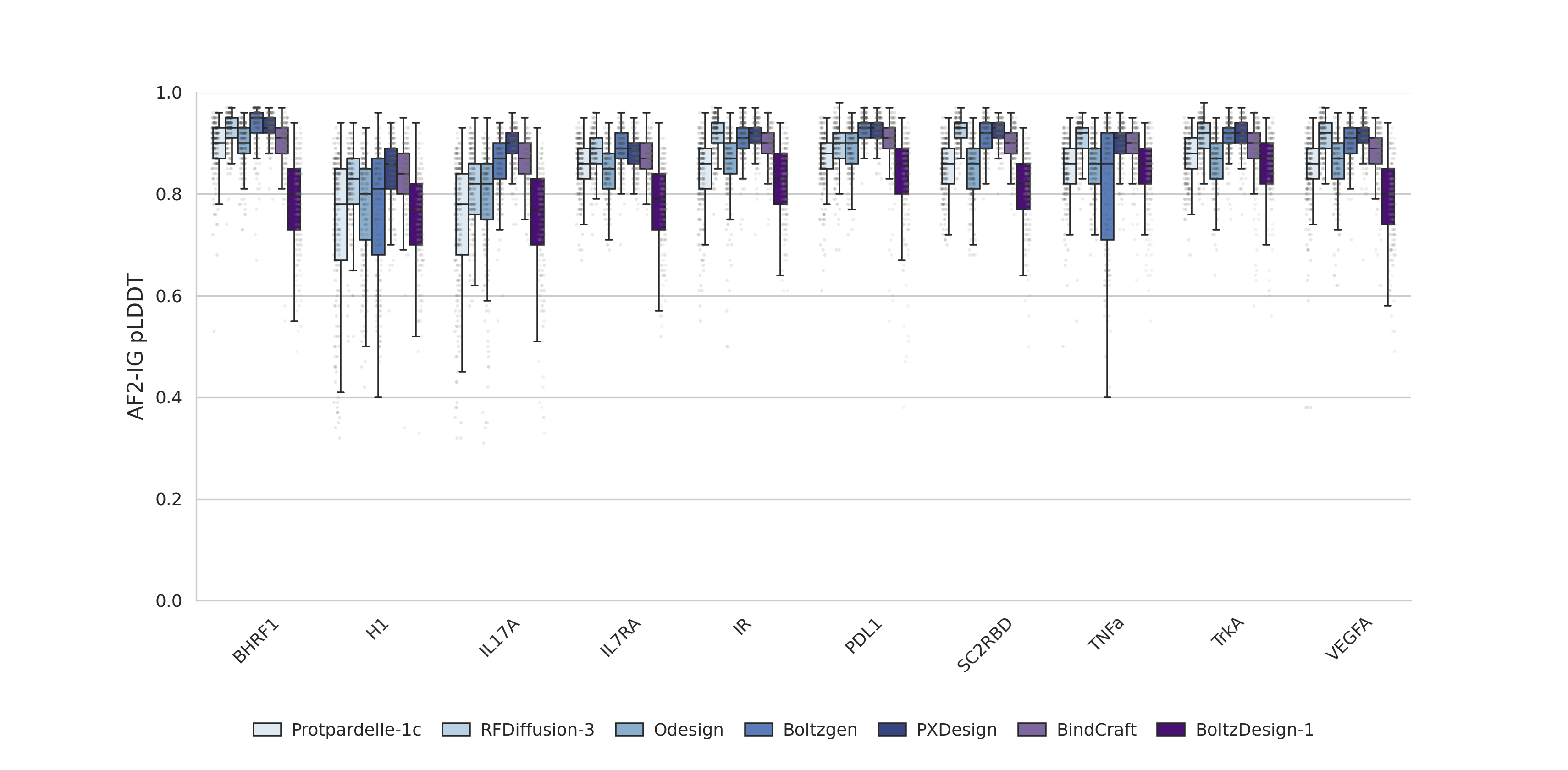}
    \caption{AlphaFold2 initial guess pLDDT on different targets across various methods.}
    \label{fig:benchmark_plddt}
\end{figure}

\begin{figure}[htbp]
    \centering
    \includegraphics[width=0.90\textwidth]{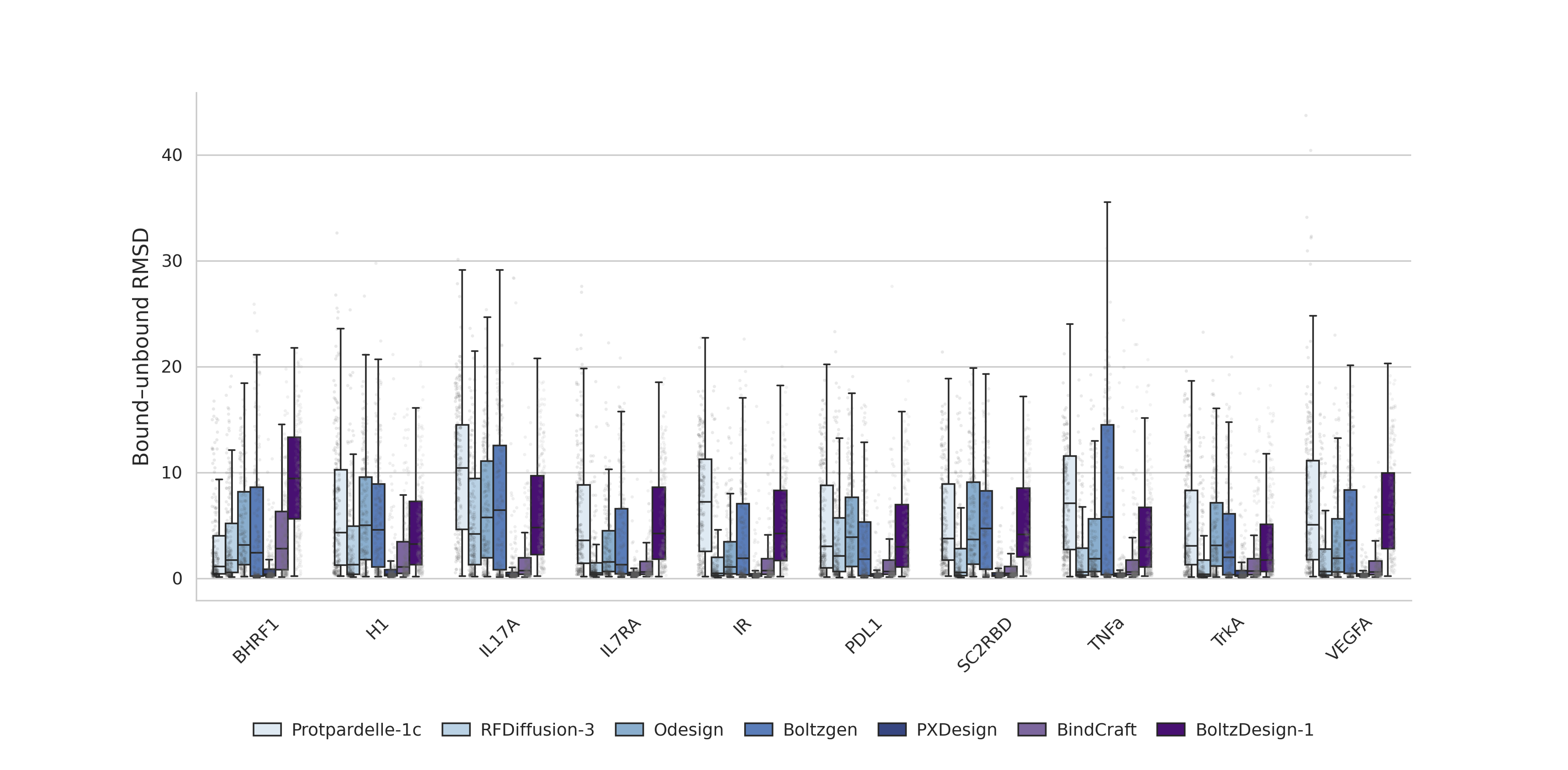}
    \caption{Bound unbound RMSD on different targets across various methods.}
    \label{fig:benchmark_bound_unbound}
\end{figure}

\begin{figure}[htbp]
    \centering
    \includegraphics[width=0.90\textwidth]{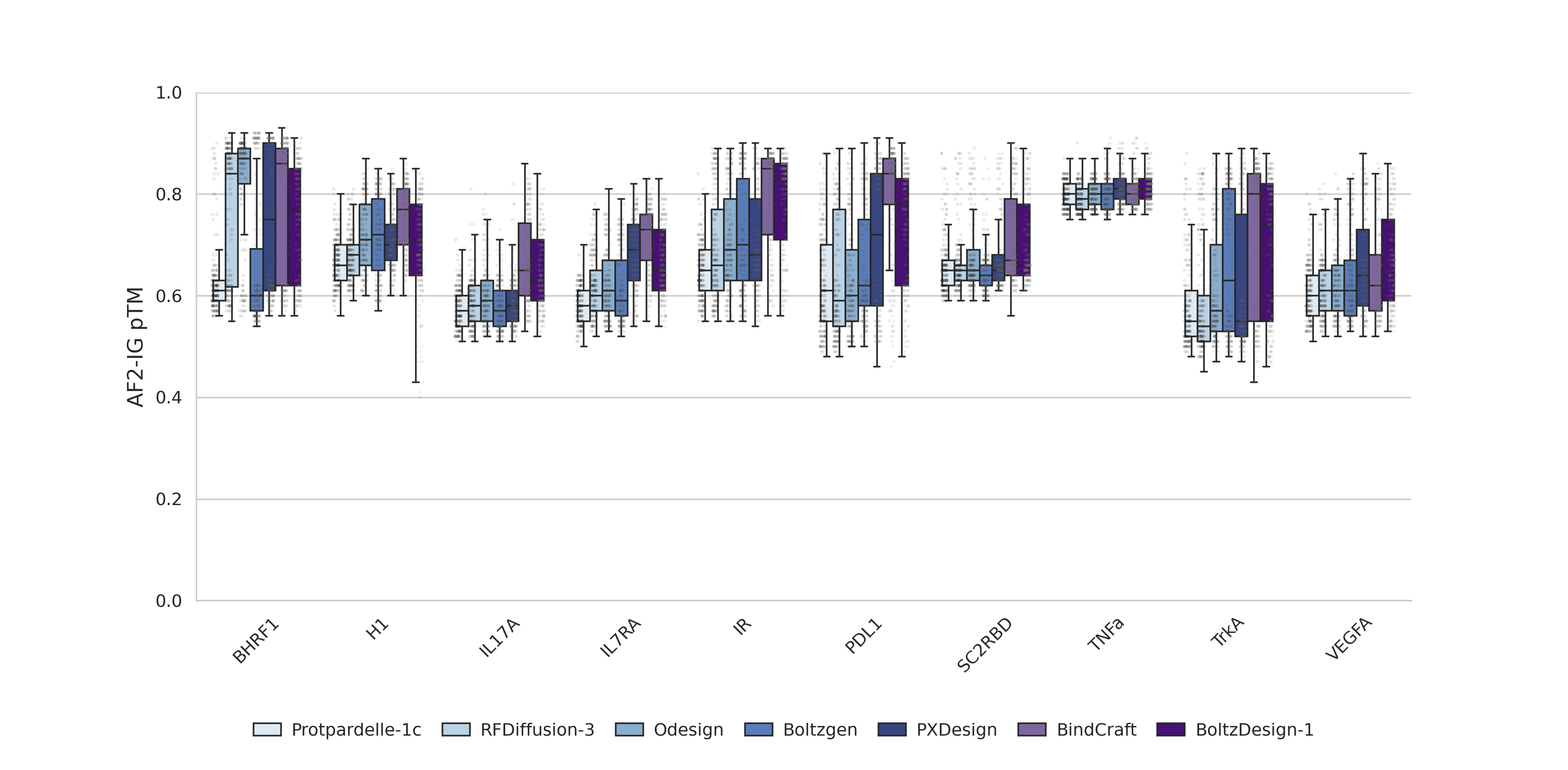}
    \caption{AlphaFold2 initial guess pTM on different targets across various methods.}
    \label{fig:benchmark_af2_ptm}
\end{figure}





\subsection{Supplementary Results on Additional Cao Targets}
\label{appendix:supp-cao-targets}

To complement the ten benchmark targets used in \Cref{sec:gen_benchmark}, we additionally report per-method results on five Cao targets (EGFR, TGF$\beta$, PDGFR, FGFR2, VirB8) under the same AF2-IG-Easy filter and ProteinMPNN sequence-design protocol.
These are not part of the main benchmark target set, but they overlap with the Cao retrospective verifier analysis (\Cref{sec:filter}) and provide a cross-check that the relative trends across methods are consistent on a different target distribution where verifier behavior is partially characterized.
\Cref{tab:supp-cao-seq} reports per-sequence AF2-IG-Easy success rate (\%), and \Cref{tab:supp-cao-cluster} reports the corresponding diversity-adjusted cluster pass rate at TM-score threshold $0.6$, matching the granularity used for \Cref{fig:binder_diversity} in the main text.

\begin{table}[h]
\centering
\small
\setlength{\tabcolsep}{6pt}
\caption{\textbf{Per-sequence AF2-IG-Easy success rate (\%) on five additional Cao targets.}
Same evaluation protocol as \Cref{fig:combined-large}a but on the five additional targets used in our retrospective verifier analysis.}
\label{tab:supp-cao-seq}
\begin{tabular}{l|rrrrr}
\toprule
\textbf{Method} & \textbf{EGFR} & \textbf{TGF}$\boldsymbol{\beta}$ & \textbf{PDGFR} & \textbf{FGFR2} & \textbf{VirB8} \\
\midrule
Protpardelle-1c & 0.04  & 0.48  & 0.00  & 1.32  & 7.70  \\
RFDiffusion-3   & 0.18  & 0.88  & 26.94 & 0.70  & 39.00 \\
ODesign         & 0.09  & 10.30 & 9.82  & 0.09  & 0.35  \\
BoltzGen        & 0.13  & 9.90  & 34.15 & 0.62  & 3.74  \\
PXDesign        & 0.13  & 8.10  & 29.80 & 4.45  & 45.82 \\
BindCraft       & 9.88  & 26.81 & 54.19 & 3.84  & 16.04 \\
BoltzDesign-1   & 2.95  & 8.89  & 45.73 & 10.04 & 2.20  \\
\bottomrule
\end{tabular}
\end{table}

\begin{table}[h]
\centering
\small
\setlength{\tabcolsep}{6pt}
\caption{\textbf{Diversity-adjusted cluster pass rate (\%) at TM-score $\geq 0.6$ on five additional Cao targets.}
Same metric as \Cref{fig:binder_diversity}.}
\label{tab:supp-cao-cluster}
\begin{tabular}{l|rrrrr}
\toprule
\textbf{Method} & \textbf{EGFR} & \textbf{TGF}$\boldsymbol{\beta}$ & \textbf{PDGFR} & \textbf{FGFR2} & \textbf{VirB8} \\
\midrule
Protpardelle-1c & 0.52  & 0.52  & 0.00  & 0.52  & 0.52  \\
RFDiffusion-3   & 1.56  & 3.65  & 45.31 & 1.04  & 28.65 \\
ODesign         & 0.52  & 0.52  & 0.52  & 0.52  & 0.52  \\
BoltzGen        & 1.04  & 12.50 & 22.40 & 2.08  & 5.73  \\
PXDesign        & 0.52  & 0.52  & 0.52  & 0.52  & 0.52  \\
BindCraft       & 20.12 & 41.51 & 48.68 & 15.03 & 28.75 \\
BoltzDesign-1   & 11.46 & 14.58 & 32.29 & 19.79 & 5.21  \\
\bottomrule
\end{tabular}
\end{table}

The relative ordering of methods on these five targets is broadly consistent with the main benchmark (\Cref{fig:combined-large,fig:binder_diversity}): PXDesign and BoltzGen are competitive among diffusion-based methods on per-sequence success rate, while BindCraft attains higher per-sequence success but lower throughput.
We provide these numbers as a cross-target sanity check rather than a head-to-head ranking, since target difficulty varies substantially across the five (e.g., EGFR is uniformly hard for all methods, whereas VirB8 admits much higher success rates).

\subsection{Runtime Measurement Protocol and Breakdown}
\label{appendix:runtime}

\paragraph{Measurement protocol.}
All generation and evaluation timing results in \Cref{sample-table-af2-comprehensive} are obtained on a single (normalized) NVIDIA A100 GPU.
We keep each method's repository-default inference behavior throughout, without method-specific re-engineering, so that the benchmark reflects practical end-to-end usage of publicly available methods rather than per-method optimized deployments that would require unequal engineering effort.
For diffusion-based methods we use generation batch size $4$, since for each target and sampled length we generate $4$ binder candidates in parallel.
The total runtime $T_{\mathrm{tot}} = T_{\mathrm{sample}} + T_{\mathrm{MPNN}} + T_{\mathrm{eval}}$ in \Cref{eq:n24h} aggregates over backbone generation, ProteinMPNN inverse folding, and verifier evaluation.
$T_{\mathrm{MPNN}}$ is small ($<1\%$ of $T_{\mathrm{tot}}$) for every method we measure and is therefore folded into the ``evaluation'' bucket in the runtime breakdown of \Cref{tab:runtime-breakdown} (reported in the main text, \Cref{sec:gen_benchmark}).

\subsection{Implementation Details}
\label{appendix:implementation_details}
The GitHub repositories and commit ids for the benchmarked methods are detailed in Table \ref{table:other_methods}. With the exception of BindCraft, all methods were evaluated using their default code and parameters.

 Since BindCraft integrates hallucination and evaluation in a single pipeline, we removed evaluation time from our measurement to enable a fairer comparison. Specifically, we measured the time consumed by the 
\texttt{binder\_hallucination} function (\url{https://github.com/martinpacesa/BindCraft/blob/main/bindcraft.py}, Lines 109-111). Within this function, we counted only the time taken to hallucinate the binder, excluding the time spent on trajectory checking (\url{https://github.com/martinpacesa/BindCraft/blob/main/functions/colabdesign_utils.py}, Lines 177-233).  

\begin{table}[ht]
\centering
\small
\renewcommand{\arraystretch}{1.2}
\setlength{\tabcolsep}{4pt}
\caption{\textbf{Details on running the compared methods.}}
\label{table:other_methods}
\begin{tabular}{
  >{\raggedright\arraybackslash}p{2.5cm}
  >{\raggedright\arraybackslash}p{12.0cm}
}
\hline
\textbf{Method} & \textbf{GitHub repository and Commit}   \\
\hline
Protpardelle-1c & \url{https://github.com/ProteinDesignLab/protpardelle/tree/main}   \\
{} & \texttt{68eb81290fd126e10a50e10d94b00c811fe7245a} \\
\hline
RFdiffusion-3 & \url{https://github.com/RosettaCommons/foundry/tree/production}   \\
{} & \texttt{cc327b4ad04551cc00bbbc3e0cdde46a7a358bde} \\
\hline
ODesign & \url{https://github.com/The-Institute-for-AI-Molecular-Design/ODesign/tree/main}   \\
{} & \texttt{aacceaa0d6cee551fa9ff886bfddfb3ae2b568d1} \\
\hline
BoltzGen & \url{https://github.com/HannesStark/boltzgen/tree/main}   \\
{} & \texttt{247b9bbd8b68a60aba854c2968d6a0cddd21ad6d} \\
\hline
PXDesign & \url{https://github.com/bytedance/PXDesign/tree/main}   \\
{} & \texttt{f788441313c84c3074fe9596ac2433f96b15c763} \\
\hline
BindCraft & \url{https://github.com/martinpacesa/BindCraft/tree/main} \\
{} & \texttt{05702c435e2172a99c2b3faf87487badb6e54727} \\
\hline
BoltzDesign-1 & \url{https://github.com/yehlincho/BoltzDesign1} \\
{} & \texttt{627c0cc7bab41e56f544c5d15467b2dbeb490168} \\
\hline
\end{tabular}
\end{table}
\section{Filtering Methodology and Benchmark Evaluation}

\subsection{Cao Dataset Summary}
\label{appendix:cao-stats}

\Cref{tab:cao-stats} summarizes the nine Cao targets used in \Cref{fig:filter-benchmark}.
For each target we report the total number of experimentally annotated non-binders (\textbf{Neg.}) and binders (\textbf{Pos.}), the number of non-binders retained after subsampling (\textbf{Ret.}, capped at $20{,}000$ per crop for tractability in \Cref{fig:filter-benchmark}a), and binder-sequence length statistics before/after subsampling.
The EGFR row pools two crops, EGFRc and EGFRn, that share the same target and are jointly displayed as ``EGFR'' in \Cref{fig:filter-benchmark}; counts are summed and length statistics are weighted by non-binder count.
The amino-acid composition total-variation distance (\textbf{TV}) between the full and subsampled non-binder pools is reported in the last column; values are uniformly small ($\leq 0.0015$), indicating that the main distributional characteristics of the non-binder pool are preserved by subsampling.
In total, the nine targets contain over $478{,}000$ non-binders and $1{,}385$ wet-lab-confirmed binders.

\begin{table}[h]
\centering
\small
\setlength{\tabcolsep}{4pt}
\caption{\textbf{Statistics of the nine Cao targets used in \Cref{fig:filter-benchmark}.}
``Ret.'' denotes the number of non-binders retained after random subsampling for the single-score ranking analysis in \Cref{fig:filter-benchmark}a (capped at $20{,}000$ per crop); the combined-filter analysis in \Cref{fig:filter-benchmark}b uses the full non-binder pool.
Mean/median length columns report binder-sequence length (in residues) on the full set vs.\ on the retained subsampled set.
``TV'' is the total-variation distance between full and subsampled amino-acid composition distributions.
EGFR pools two crops (EGFRc, EGFRn).}
\label{tab:cao-stats}
\begin{tabular}{lrrrlll}
\toprule
\textbf{Target} & \textbf{Neg.} & \textbf{Pos.} & \textbf{Ret.} & \textbf{Mean len.} & \textbf{Median len.} & \textbf{TV} \\
\midrule
EGFR            & 99{,}788 & 15   & 40{,}000 & 63.28/63.27 & 65/65 & 0.0012 \\
FGFR2           & 59{,}395 & 604  & 20{,}000 & 63.54/63.57 & 65/65 & 0.0015 \\
IL-7RA          & 14{,}977 & 22   & 14{,}977 & 61.72/61.72 & 63/63 & 0.0000 \\
InsulinR        & 59{,}714 & 259  & 20{,}000 & 65.00/65.00 & 65/65 & 0.0010 \\
PDGFR           & 99{,}708 & 284  & 20{,}000 & 65.00/65.00 & 65/65 & 0.0012 \\
SARS-CoV-2 RBD  & 99{,}980 & 19   & 20{,}000 & 65.00/65.00 & 65/65 & 0.0012 \\
TGF$\beta$      & 14{,}900 & 100  & 14{,}900 & 65.00/65.00 & 65/65 & 0.0000 \\
TrkA            & 14{,}990 & 10   & 14{,}990 & 60.93/60.93 & 62/62 & 0.0000 \\
VirB8           & 14{,}939 & 72   & 14{,}939 & 63.31/63.31 & 65/65 & 0.0000 \\
\midrule
\textbf{Total} & \textbf{478{,}391} & \textbf{1{,}385} & \textbf{179{,}806} & --- & --- & --- \\
\bottomrule
\end{tabular}
\end{table}


\subsection{Filter Combination Search}
\label{appendix:filter-search}
To identify general-purpose, model-specific filters, we perform a grid search over combinations of confidence scores. 
We design a two-stage filtering strategy to identify high-quality protein complex predictions generated by Protenix. In the first stage, we identify informative evaluation metrics and select an optimal triplet combination based on their ranking performance across targets. In the second stage, we perform a grid search over the cutoff values of the selected metrics to determine optimal thresholds for filtering.

\textbf{Stage 1: Metric Selection and Combination.}
We first evaluate the ranking performance of individual metrics using a Top-1\% threshold classifier and compute the Enrichment Factor (EF) to measure how effectively each metric identifies high-quality predictions. For each target, we select the Top-3 metrics with the highest EF values. Based on frequency analysis, we identify eight most frequently selected metrics (some with the same EF score): binder pTM, complex pLDDT, interface pLDDT, binder ipTM, complex pTM, binder chain pLDDT, complex gpDE, and interface gpDE.
From these eight metrics, we evaluate all possible triplets ($C(8,3) = 56$ combinations) as filters. For each triplet, we apply the Top-1\% quantile as a threshold and classify a sample as positive only if it satisfies all three thresholds simultaneously. We compute the Success Rate (SR) for each combination on every target and rank them accordingly. The final optimal combination is determined by aggregating rankings across all targets and selecting the one that achieves the most ``Top-1'' positions. The final optimal combination is determined by aggregating rankings across all targets and selecting the one that achieves the highest ranking among all targets. Ultimately, we find that a simplified filter using only two metrics, binder pTM and binder ipTM, achieves comparable or better performance than the full triplet, while offering improved robustness and interpretability. Therefore, we adopt $\{\text{binder pTM}, \text{binder ipTM}\}$ as the final filtering metric set.

\textbf{Stage 2: Cutoff Grid Search.}
Finally, to refine the filtering process, we perform a grid search to determine the optimal cutoff values for the two selected metrics, binder pTM and binder ipTM.
We formulate the filter selection problem as an optimization problem, 
\begin{equation}
\begin{aligned}
    \underset{\mathbf{x} \in \mathcal{X}}{\max}  \left( \textrm{SR}_1(\mathbf{x}), \textrm{SR}_2(\mathbf{x}), \ldots, \textrm{SR}_k(\mathbf{x}) \right),
\end{aligned}
\end{equation}
in which $\textrm{SR}_i$ denotes the success rate on the target $i$, and the set $\mathcal{X}$ is the feasible set of confidence score threshold combination. 
Typically, there is no feasible solution that can maximize all objective functions simultaneously. Consequently, the focus is the solutions where improving any objective cannot be achieved without deteriorating at least one other objective, which is defined as \emph{Pareto Frontier}.

\emph{Definition:} A solution $x_1 \in \mathcal{X}$ dominates $x_2$, if 
\begin{equation}
    \forall i, \textrm{SR}_i(x_1) \geq \textrm{SR}_i(x_2);
    \exists i, \textrm{SR}_i(x_1) > \textrm{SR}_i(x_2).
\end{equation}
A solution $x^* \in \mathcal{X}$ is Pareto optimal if there does not exist another solution $x$ that dominates it. The set of Pareto optimal is called Pareto Frontier. 
%

\begin{figure}[h]
\centering
\setlength{\tabcolsep}{3pt}
\begin{tabular}{cccc}
\raisebox{1.4em}{\rotatebox{90}{\tiny IL-7RA SR}}
\includegraphics[width=0.215\textwidth]{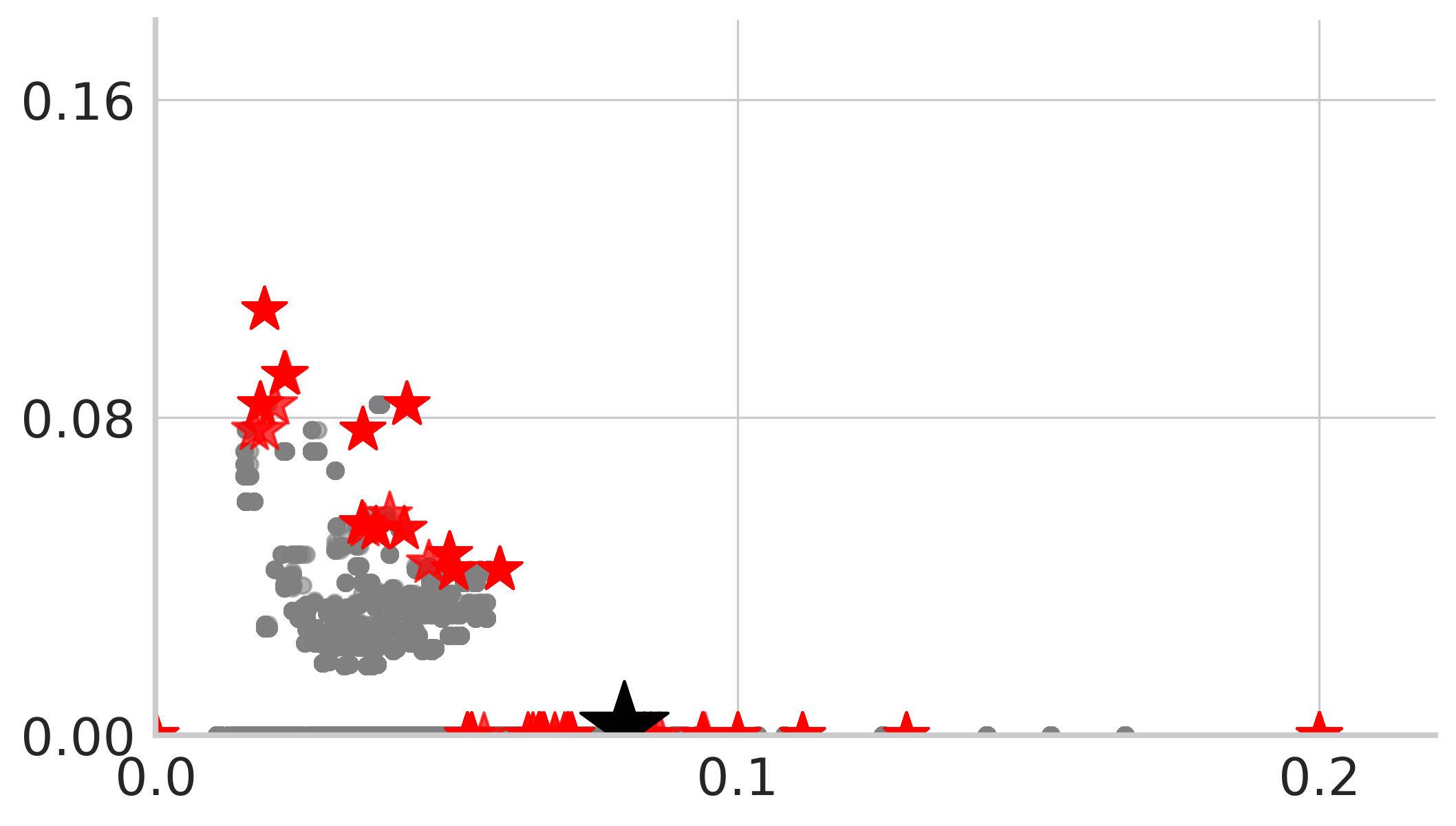}
& 
\raisebox{1.1em}{\rotatebox{90}{\tiny Insulin SR}}
\includegraphics[width=0.215\textwidth]{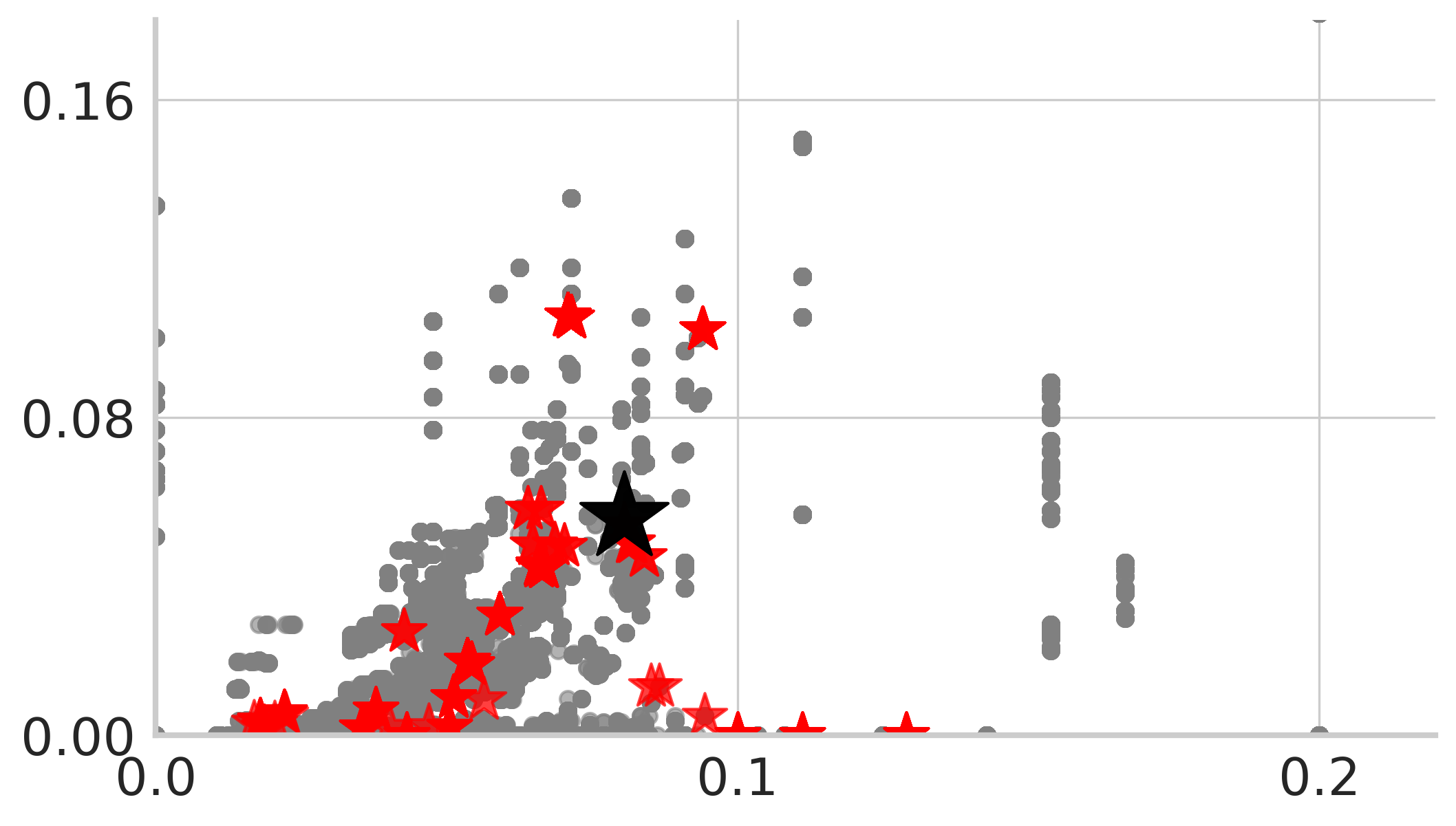}
& 
\raisebox{0.8em}{\rotatebox{90}{\tiny SC2RBD SR}} \includegraphics[width=0.215\textwidth]{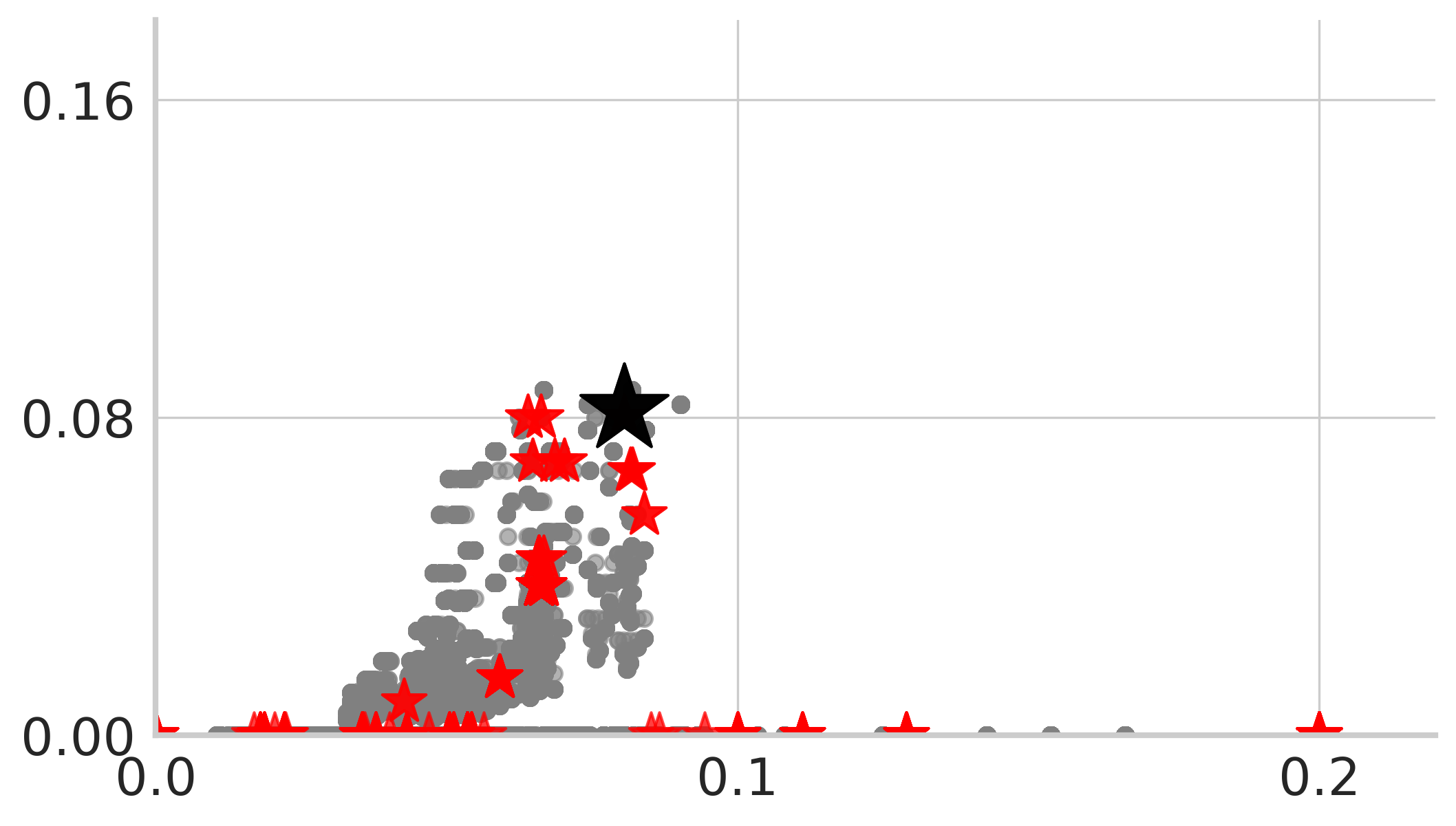}
& 
\raisebox{1.2em}{\rotatebox{90}{\tiny TGFb SR}}
\includegraphics[width=0.215\textwidth]{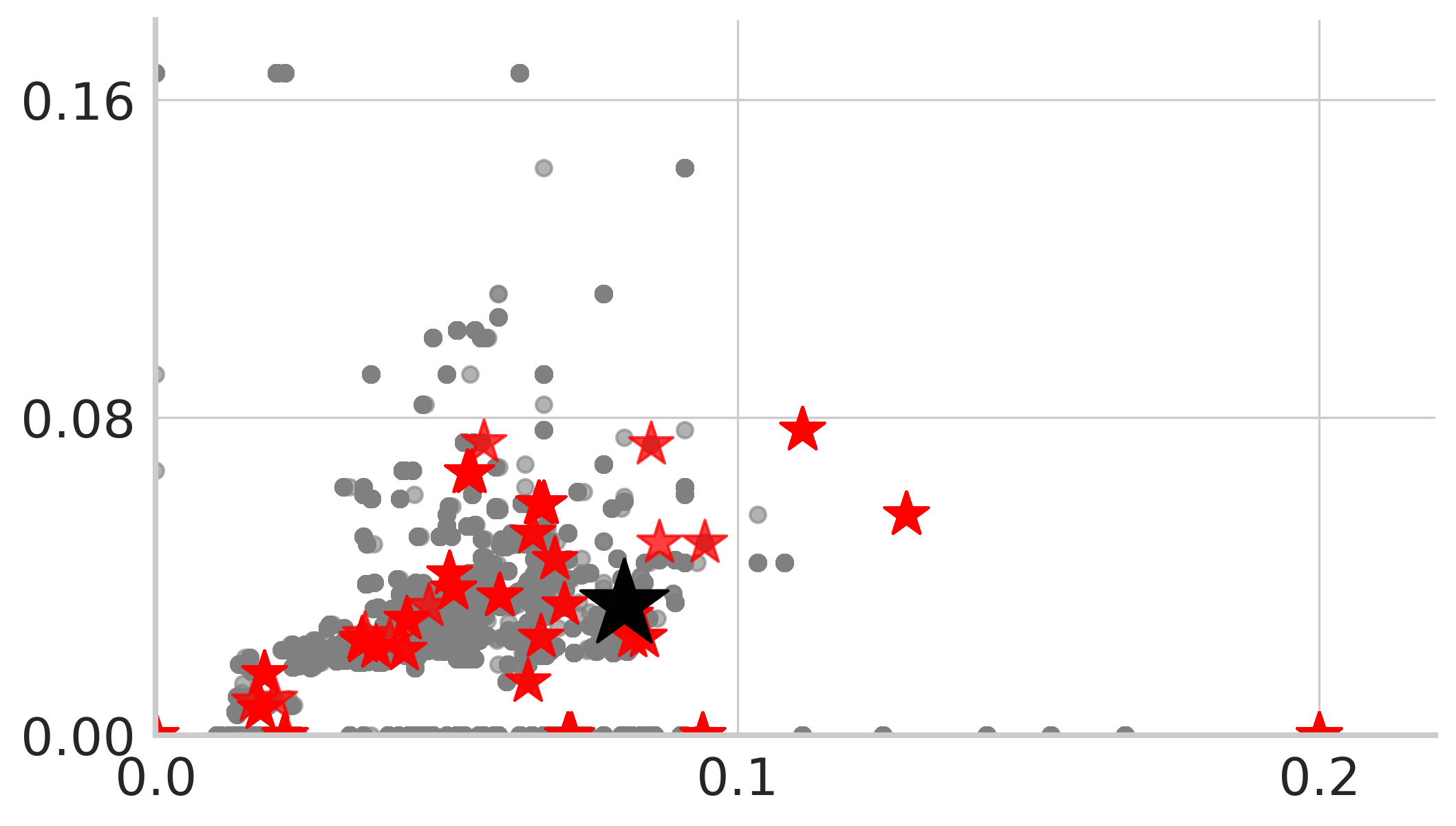}
\\
{\tiny FGFR2 SR} & {\tiny FGFR2 SR}  & {\tiny FGFR2 SR}  & {\tiny FGFR2 SR}  \\

\end{tabular}
\caption{\textbf{The performance of the AF2 confidence score filters.} The SR for each confidence combination is plotted as one gray dot. The pareto frontier filters are highlighted as red stars, and the selected one is marked as a black star.} 
\label{fig:af2-pareto}
\end{figure}

For each threshold combination, we compute the success rate (SR) on each target. 
Following the definition of Pareto Frontier, the search algorithm can come to a set of optimal points (as demonstrated in \Cref{fig:af2-pareto}).
To distinguish the final solution, we tend to the solution which has minimal shifts across different targets, known as robust selection or risk aversion policy \citep{asimit2017robust}. 
We calculate the rank of each combination's SR within each target and then take the average rank across all targets. This average rank serves as the overall ``score'' for the threshold combination. As demonstrated in the third figure in \Cref{fig:af2-pareto}, we come to a balanced SR solution on FGFR2 and SC2RBD.
Ultimately, we select the combination with the highest score as the final filtering criteria.

\subsection{Per-Target Agreement Curves}
\label{appendix:agreement-curves}

\begin{figure}[h!]
\centering
\includegraphics[width=\linewidth]{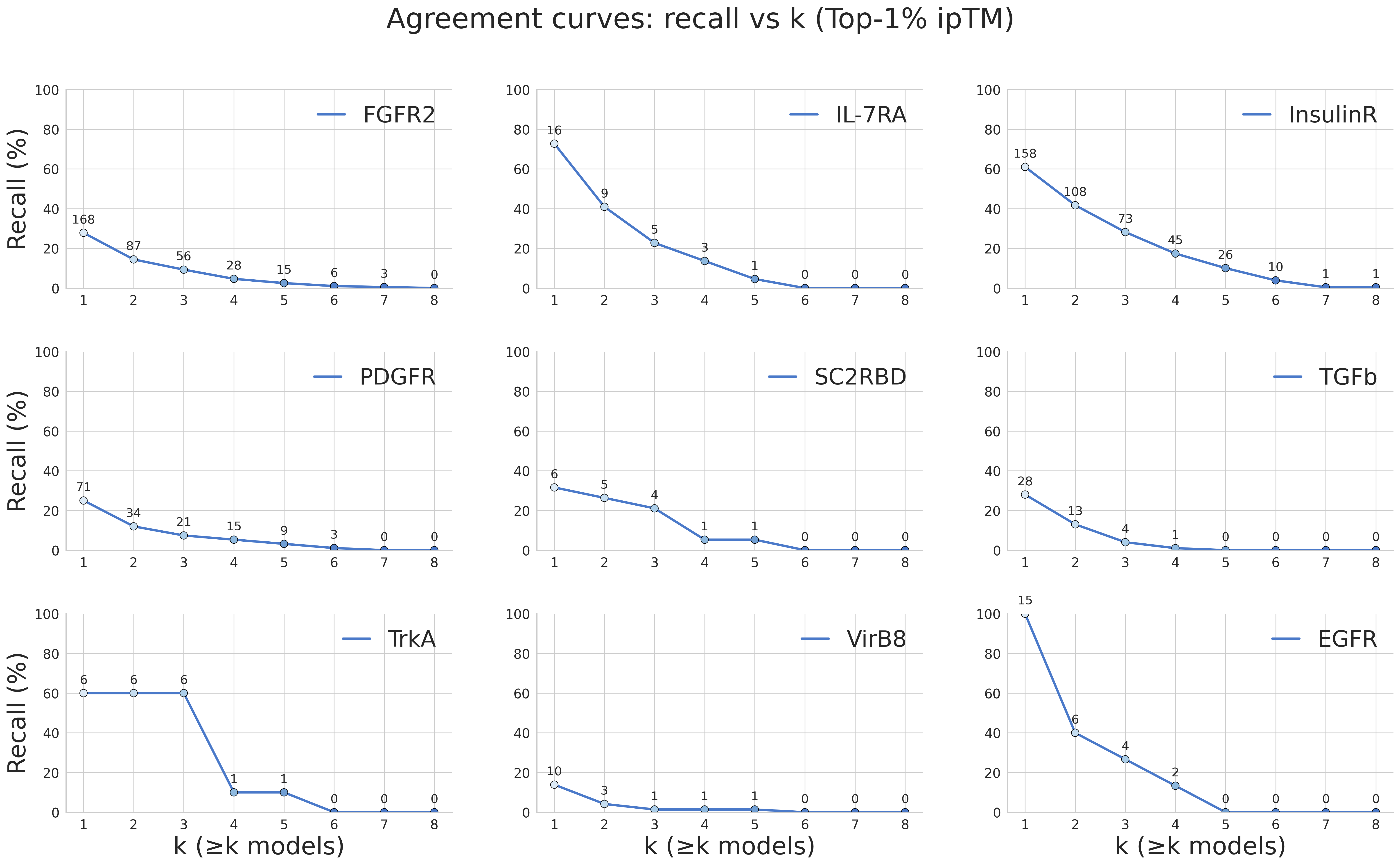}
\caption{\textbf{Per-target agreement curves under Top-1\% ipTM filtering.}
For each target, recall (\%) is plotted against the agreement threshold $k$ (a true positive must be identified by at least $k$ verifiers simultaneously).
Annotations on each curve indicate the number of true positives that satisfy the agreement threshold.
Recall decreases rapidly as $k$ increases, indicating limited consensus among verifiers and highlighting their complementary coverage of the binder design space.}
\label{fig:agreement-curves}
\end{figure}

\Cref{fig:filter-benchmark}c reports the union recall under Top-1\% ipTM filtering aggregated to one bar per target.
\Cref{fig:agreement-curves} provides the per-target breakdown: recall as a function of the agreement threshold $k$, where a true positive must be identified by at least $k$ of the verifiers simultaneously.

\subsection{Per-Score Filter Evaluation}
\label{appendix:filter-metrics}

To complement the Top-1\% SR analysis in the main text (\Cref{fig:filter-benchmark}a), we provide additional evaluation of individual confidence scores using two standard ranking metrics: AUC (area under the ROC curve) and AP (average precision, or area under the PR curve), evaluated on the same nine subsampled Cao targets used in \Cref{fig:filter-benchmark}a (FGFR2, IL-7RA, InsulinR, PDGFR, SC2RBD, TGF$\beta$, TrkA, VirB8, EGFR). These metrics reflect how well each score discriminates binders from non-binders across a range of thresholds, independent of any fixed selection cutoff.

\begin{figure}[h!]
\centering

    \begin{minipage}[t]{0.99\textwidth}
        \raisebox{8pt}{\textbf{(a)}}\\[-1.5ex]
        \includegraphics[width=\linewidth]{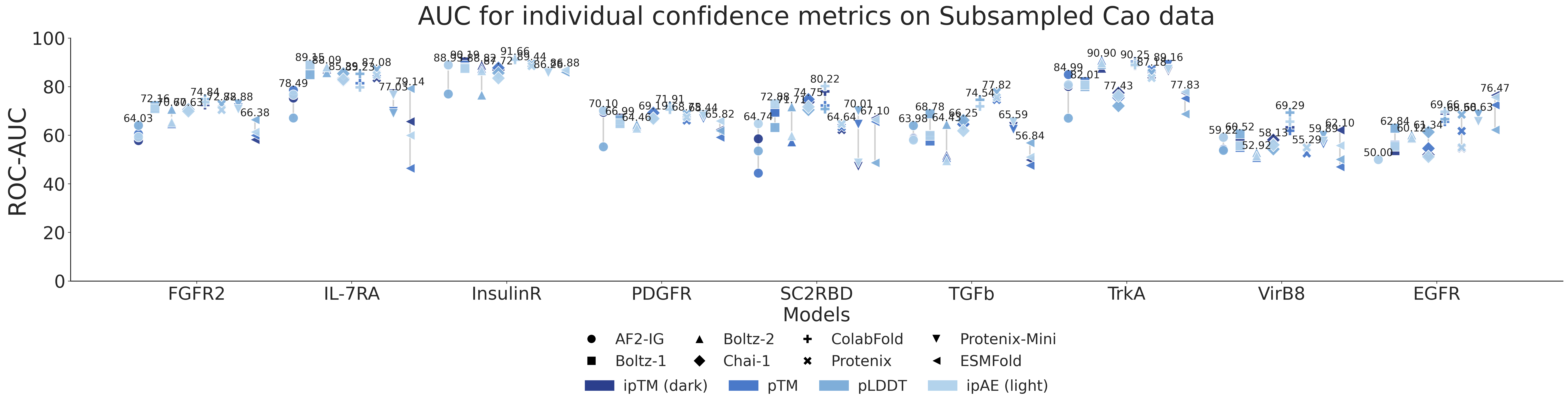}
    \end{minipage}

    \vskip 1em

    \begin{minipage}[t]{0.99\textwidth}
        \raisebox{8pt}{\textbf{(b)}}\\[-1.5ex]
        \includegraphics[width=\linewidth]{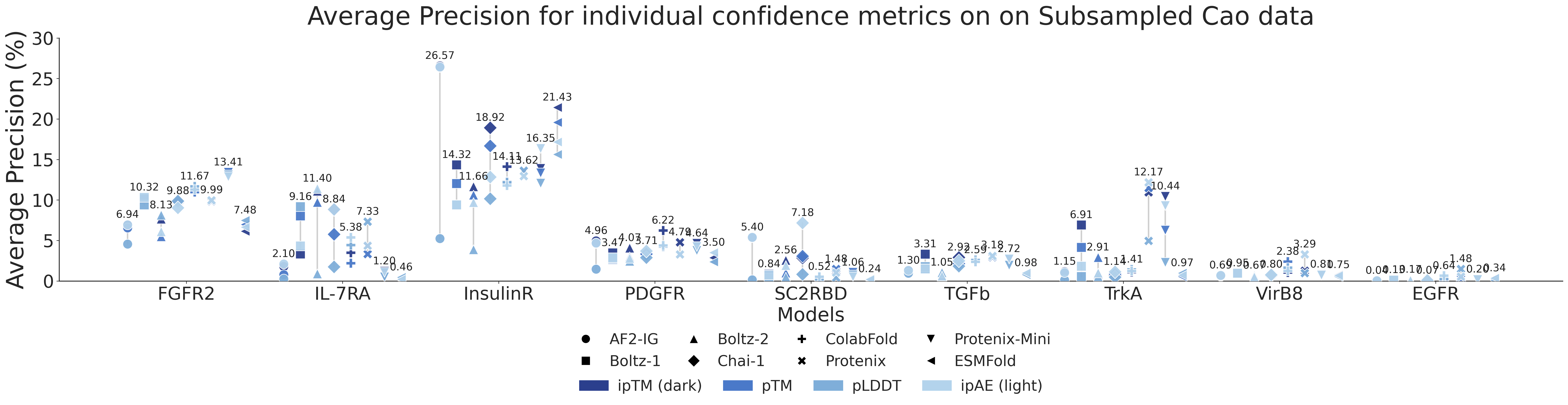}
    \end{minipage}
\caption{\textbf{AUC and Average Precision scores for individual confidence metrics on the nine subsampled Cao targets.}
Same target set and verifier set as \Cref{fig:filter-benchmark}a.
\textbf{(a)} ROC-AUC: higher values indicate better global discrimination between binders and non-binders, integrated across all decision thresholds.
\textbf{(b)} Average Precision (area under the PR curve): more sensitive than AUC to top-ranking false positives, and therefore more aligned with the Top-1\% screening regime in \Cref{fig:filter-benchmark}a.
}
\label{fig:per-score-auc-ap}
\end{figure}

We report AUC and average precision scores for individual confidence metrics across diverse design targets (\Cref{fig:per-score-auc-ap}). These results are generally consistent with the Top-1\% SR trends, reinforcing that Protenix-derived scores perform competitively to other verifiers. However, no single metric is universally optimal—performance varies by target and model.





\subsection{Retrospective Evaluation on AI-generated RFdiffusion Designs}
\label{appendix:rfdiffusion-retrospective}

As a complementary study, we evaluate two representative verifiers from \Cref{tab:filter-thresholds}, AF2-IG and Protenix-Mini, against the wet-lab outcomes of binders released with RFdiffusion~\citep{watson2023novo}.
Per-target results are reported in \Cref{tab:rfdiff-retrospective}.
On this AF2-IG-pre-filtered AI-generated dataset, Protenix-Mini is more selective and retains fewer candidates, but achieves non-trivial precision on Mdm2 ($0.623$), PDL1 ($0.389$), IL7Ra ($0.414$), and InsulinReceptor ($0.300$), suggesting that our framework remains meaningfully correlated with wet-lab outcomes on AI-generated candidate distributions and not only on the unfiltered Cao setting.

\begin{table}[h]
\centering
\small
\setlength{\tabcolsep}{4pt}
\caption{\textbf{Retrospective verifier evaluation on AF2-IG-prefiltered RFdiffusion designs.}
``Wet+/Wet$-$'' are the wet-lab confirmed binder/non-binder counts per target. ``TP'' is the number of true positives retained by each verifier; precision, recall, and $F_1$ are computed against wet-lab labels. The last row aggregates across all six targets.}
\label{tab:rfdiff-retrospective}
\begin{tabular}{lrr|rrrr|rrrr}
\toprule
& & & \multicolumn{4}{c|}{\textbf{AF2-IG}} & \multicolumn{4}{c}{\textbf{Protenix-Mini}} \\
\textbf{Target} & \textbf{Wet+} & \textbf{Wet$-$} & TP & Prec.\ & Rec.\ & $F_1$ & TP & Prec.\ & Rec.\ & $F_1$ \\
\midrule
IL7Ra            & 32  & 63  & 23  & 0.324 & 0.742 & 0.451 & 12 & 0.414 & 0.387 & 0.400 \\
InfluenzaHA      & 16  & 79  & 1   & 0.125 & 0.062 & 0.083 & 0  & 0.000 & 0.000 & 0.000 \\
InsulinReceptor  & 18  & 77  & 14  & 0.173 & 0.778 & 0.283 & 6  & 0.300 & 0.333 & 0.316 \\
Mdm2             & 55  & 41  & 50  & 0.588 & 0.909 & 0.714 & 33 & 0.623 & 0.600 & 0.611 \\
PDL1             & 12  & 83  & 10  & 0.128 & 0.833 & 0.222 & 7  & 0.389 & 0.583 & 0.467 \\
TrkA             & 6   & 89  & 6   & 0.067 & 1.000 & 0.126 & 0  & 0.000 & 0.000 & 0.000 \\
\midrule
\textbf{All}     & \textbf{139} & \textbf{432} & \textbf{104} & \textbf{0.252} & \textbf{0.754} & \textbf{0.378} & \textbf{58} & \textbf{0.479} & \textbf{0.420} & \textbf{0.448} \\
\bottomrule
\end{tabular}
\end{table}

\section{Related Work}

\paragraph{Protein binder design.}

Previous works have exemplified binder design for some targets using natural interaction motifs or by generating or retrieving motifs computationally~\citep{cao2022design}, and subsequently grafting them onto pre-defined scaffold libraries. These designs are then refined through sequence optimization to improve motif–scaffold compatibility and binder–target affinity using tools such as RosettaDesign~\citep{kortemme2002simple}. However, the experimental success rate remains low, and thus these approaches often rely on high-throughput experimental techniques to achieve functional binders. Moreover, the affinity of directly designed binders is typically insufficient, often requiring further rounds of experimental affinity maturation to reach nanomolar levels. Recent advances enable \textit{de novo} binder generation via deep generative models~\citep{watson2023novo,krishna2024generalized,biogeometry2025geoflow,chai2025zero,bridgland2025latent,pacesa2025one, ren2025pxdesign, zhang2025odesignworldmodelbiomolecular, Stark2025BoltzGen}. These models are typically evaluated \textit{in silico} on different targets using structure prediction models confidence scores (e.g., pLDDT, ipAE) and self-consistency RMSD. However, evaluation protocols vary significantly across studies. As a result, it remains unclear how reported performance differences reflect underlying model capabilities versus evaluation choices.

\paragraph{Structure prediction models.}
Modern binder design relies heavily on accurate structure prediction. Tools like trRosetta~\cite{yang2020improved}, RoseTTAFold~\cite{baek2021accurate}, and AlphaFold2~\cite{jumper2021highly} enable accurate prediction of protein monomer structure. More recent models such as ESMFold~\cite{lin2023evolutionary} and AlphaFold-Multimer~\cite{evans2021protein} extend these capabilities to protein multimers. AlphaFold3(AF3)~\citep{abramson2024accurate} achieves higher prediction accuracy and is able to predict the joint structure
of complexes including proteins, nucleic acids, small molecules, ions and modified residues. Multiple open-source variants of AF3 are released later, including Boltz-1~\citep{wohlwend2024boltz}, Boltz-2~\citep{passaro2025boltz2}, Chai-1~\citep{chai2024chai} and Protenix~\citep{chen2025protenix}. 
These predictors are now integral to design and evaluation workflows.

\end{document}